\documentclass[12pt]{iopart}

\usepackage{amsfonts}
\usepackage{amssymb}
\usepackage{amsbsy}
\usepackage{graphicx}
\usepackage{algorithm}
\usepackage[noend]{algpseudocode}
\usepackage{comment}
\usepackage{hyperref}
\usepackage{color}
\usepackage{lineno}
\usepackage{ulem}
\usepackage{siunitx}

\graphicspath{ {fig/}}

\newcommand{\abs}[1]{
	|#1|
}

\begin{document}
    \title{Source term method for binary neutron stars initial data}

    \author{Bing-Jyun Tsao}
    \address{National Center for Supercomputing Applications, 
             University of Illinois at Urbana-Champaign, Urbana, Illinois 61801, USA.}
    \ead{btsao2@illinois.edu}
    
    \author{Roland Haas}
    \address{National Center for Supercomputing Applications, 
             University of Illinois at Urbana-Champaign, Urbana, Illinois 61801, USA.}
    
    \author{Antonios Tsokaros}
    \address{Department of Physics, University of Illinois at Urbana-Champaign, Urbana, Illinois 61801, USA.}

    \begin{abstract}
    The initial condition problem for a binary neutron star system requires a Poisson equation solver for the velocity potential with a Neumann-like boundary condition on the surface of the star. 
    Difficulties that arise in this boundary value problem are: a) the boundary is not known a-priori, but constitutes
    part of the solution of the problem; b) various terms become singular at the boundary.
    In this work, we present a new method to solve the fluid Poisson equation for irrotational/spinning binary neutron
    stars. The advantage of the new method is that it does not require complex fluid surface fitted coordinates and 
    it can be implemented in a Cartesian grid, which is a standard choice in numerical relativity calculations.
    This is accomplished by employing the source term method proposed by Towers, where the boundary condition is treated as a jump condition and is incorporated as additional source terms in 
    the Poisson equation, which is then solved iteratively.
    The issue of singular terms caused by vanishing density on the surface is resolved with an additional separation that shifts the computation boundary 
    to the interior of the star.
    We present two-dimensional tests to show the convergence of the source term method, and we further apply this 
    solver to a realistic three-dimensional binary neutron star problem. By comparing our solution with the one coming 
    from the initial data solver \textsc{cocal}, we demonstrate agreement to approximately $1\%$. Our method 
    can be used in other problems with non-smooth solutions like in magnetized neutron stars.
    \end{abstract}
    
    \vspace{2pc}


\section{Introduction}
\label{sec:intro}
A system of two neutron stars (NSs) in a quasicircular orbit constitutes one of the finest laboratories in modern astrophysics since it is the place were a trinity of important phenomena intersect: a) densities of matter beyond nuclear density, b) the nucleosynthesis of heavy elements, c) the formation of powerful gamma-ray bursts. Behind the scenes, it is the synergy of all the forces in Nature with maximal strength that is responsible for the creation of the extreme conditions which lead to this intersection. Event GW170817 revealed in the most dramatic way the above crossroads and initiated the era of gravitational wave multimessenger astronomy by Advanced LIGO/Virgo~\cite{2017PhRvL.119p1101A}, the Fermi Gamma-Ray Burst Monitor~\cite{2017GCN.21520....1V,2017GCN.21517....1K} and INTEGRAL~\cite{Savchenko:2017ffs,Savchenko17GCN}. This event was the first time both the gravitational waves and the electromagnetic
signal emitted during a quasicircular binary neutron star (BNS) collision were detected \cite{TheLIGOScientific:2017qsa,GBM:2017lvd,Monitor:2017mdv,Chornock:2017sdf,2017GCN.21520....1V,Savchenko:2017ffs}.
Although the possibility of exotic compact objects or even a neutron star-black hole system could not be ruled out, the characteristics of GW170817 suggest that a BNS is the most probable scenario~\cite{TheLIGOScientific:2017qsa}.

In order to predict the outcome of a BNS merger, one needs to solve at least the Einstein and Euler equations (for more realistic scenarios additional complexity is present through Maxwell equations and many others) for two NSs in circular orbits. Initial data for such configurations were first presented by Baumgarte \textit{et al.}~\cite{1997PhRvL..79.1182B,1998PhRvD..57.7299B} and Marronetti \textit{et al.}~\cite{1998PhRvD..58j7503M}, which described two NSs in synchronous 
rotation (an example of synchronous rotation is the Moon in the Earth-Moon system). These solutions where both objects are in synchronous rotation are called corotating solutions, and they gave valuable insights of the two body problem in general relativity. Since the works of Bildsten \& Cutler~\cite{1992ApJ...400..175B} and Kochanek~\cite{1992ApJ...398..234K}, such NS configurations were considered unrealistic as viscosity is considered to be too small in order for synchronization to manifest. Despite this, corotating solutions still 
play an important role since they represent the simplest spinning binary NSs. According to our current understanding BNSs are considered to be mostly irrotational, i.e. having zero vorticity. Data describing such systems are more complicated to be computed since the conservation of rest mass is not identically satisfied as in the corotating case but results in an additional potential equation. The first numerical solutions of irrotational BNSs were presented by Bonazzola \textit{et al.}~\cite{1999PhRvL..82..892B}, Gourgoulhon \textit{et al.}~\cite{2001PhRvD..63f4029G}, Marronetti \textit{et al.}~\cite{2000NuPhS..80C0714M,1999PhRvD..60h7301M}, and Ury\=u \textit{et al.}~\cite{2000PhRvD..61l4023U,2000PhRvD..62j4015U}. Notwithstanding the fact that irrotational BNSs are expected to be the most frequent in Nature,
one cannot exclude the possibility of BNSs that exhibit spin. After all, even event GW170817 \cite{2017PhRvL.119p1101A} was unable to exclude a highly spinning binary. Solving the Euler equations in the presence of spin is a challenging problem since no trivial integral exists. There were many attempts to address that problem, for example, by Marronetti and Shapiro~\cite{Marronetti:2003gk}, Baumgarte and Shapiro~\cite{2009PhRvD..80f4009B,2009PhRvD..80h9901B}, and Tsatsin and Marronetti~\cite{2013PhRvD..88f4060T}, but the first self-consistent formulation was presented by Tichy~\cite{Tichy:2011gw,2012PhRvD..86f4024T,Tichy:2019ouu}. 

One common characteristic of both irrotational and spinning solutions is the potential equation that results from the conservation of rest mass and the symmetry of the quasicircular motion \cite{Tsokaros_2015}. In particular, the fluid velocity can be decomposed as a gradient of a potential plus a spinning part which is zero for the irrotational case. Then the divergence of the fluid velocity leads to an elliptic equation for the fluid potential with boundary conditions on the surface of the star. There are two problems that arise in this boundary value problem (BVP): a) the surface
of the star is not known and constitutes part of the solution of the problem, b) terms that contain the density of the star become singular as we approach the surface. In order to address these issues, various techniques have been employed over the past decades 
\cite{1999PhRvL..82..892B,2000PhRvD..61l4023U,Tsokaros_2015} that, despite exhibiting slow convergence, proved robust in providing
physically acceptable solutions.
    
In this paper, we present a new method to solve the nonlinear Poisson equation for the irrotational/spinning
flow in a binary neutron star system. One of the main advantages of our method is its simplicity since it does 
not require complex surface fitted coordinates and it can be implemented in a Cartesian grid which is a standard 
choice in numerical relativity calculations. This is accomplished by employing the source term method proposed by Towers~\cite{STM_2018} where the boundary condition is treated as a jump condition and is incorporated as additional 
source terms in the Poisson equation, which is then solved iteratively. As a result, the fluid sector of a binary
neutron star can be treated in similar manner as the gravitational elliptic equations with no need of extra
steps (see \cite{Tsokaros_2015}) that make the iteration process complex. The achieved generality will be very 
useful in the treatment of more complicated physical systems that have non-smooth solutions across the surface, such as
magnetized rotating neutron stars.

This article is presented in four parts. In section~\ref{sec:phys_theory}, the fluid equation for the BNS initial value problem is presented; section~\ref{sec:stm} gives a detailed introduction of the source term method that will be used for solving the Poisson problem; section~\ref{sec:numerical_discretization} describes the discretization scheme; numerical tests and comparisons with realistic initial data from the \textsc{cocal} solver are presented in section~\ref{sec:nr}. Throughout the article we use units of $G=c=M_\odot=1$, and a space-time signature $(-+++)$. Spacetime indices will be indicated with Greek letters, and spatial indices with Latin letters.

\section{Fluid equations in binary neutron stars}
\label{sec:phys_theory}
    
In this section we introduce the fluid equation describing irrotational and spinning binary neutron stars \cite{1999PhRvL..82..892B,Teukolsky98,Shibata98,Tichy:2011gw}. 
Our discussion will follow closely the works of~\cite{2000PhRvD..61l4023U,Tsokaros_2015,Tsokaros:2018dqs} where 
details are provided in full. The numerical implementation of this boundary value problem in the context of the \textsc{cocal} code is described in \cite{Uryu:2011ky,Tsokaros_2015} and will not be 
repeated here. Other successful initial data solver implementations include the \textsc{lorene} code \cite{lorene_web}, the \textsc{sgrid} code \cite{Tichy:2009yr}, and the \textsc{spells} code \cite{Foucart:2008qt}. 

\subsection*{3+1 decomposition}
\label{sec:3p1}

We assume that the spacetime $\cal M$ is foliated by a family of spacelike hypersurfaces 
$\Sigma_t$, i.e. ${\cal M} = {\mathbb R} \times \Sigma_t$ with $t\in {\mathbb R}$.  
The spacetime metric can then be written in 3+1 form as
\begin{equation}
    ds^2 = -\alpha^2 dt^2 + \gamma_{ij} (dx^i + \beta^i dt)(dx^j + \beta^j dt),
\end{equation}
where $\alpha, \beta^i, \gamma_{ij}$ are, respectively, the lapse, the shift vector, and the spatial 
three-metric on $\Sigma_t$. For the three-geometry we assume it is conformally flat 
\begin{equation}
    \gamma_{ij} = \psi^4\delta_{ij} ,
\end{equation}
where $\psi$ is the conformal factor. 

If we neglect the loss of energy due to gravitational radiation and assume
closed orbits for the BNS system, this implies the existence of a
helical Killing vector
\begin{equation} 
k^\mu \equiv t^\mu + \Omega\phi^\mu\,,
\label{eq:hkv}
\end{equation}
such that $ \mathcal{L}_{\boldsymbol k}g_{\alpha\beta}=0$, where $\mathcal{L}_{\boldsymbol k}$ is the Lie derivative along ${\boldsymbol k}$. In Eq. (\ref{eq:hkv})
$t^\mu$ is the vector that represents the flow of time, $\Omega$ is the orbital angular velocity,
and $\phi^\mu$ is the generator of rotational symmetry. In a Cartesian coordinate system, without loss of generality, we can assume $\phi^\mu=(0,\phi^i)=(0,-y,x,0)$, 
i.e. that the orbital angular momentum is along the z-axis.   
By introducing the unit timelike normal $n_\mu \equiv -\alpha \nabla_\mu t$ (where $\nabla$ denotes the covariant derivative associated with the metric $g_{\alpha\beta}$) to the hypersurfaces $\Sigma_t$, one can rewrite the Killing vector 
$k^\alpha$ as
\begin{equation}
    k^\alpha \equiv \alpha n^\mu + \omega^\mu
    \label{eq:hkv2}
\end{equation}
where $\omega^\mu = \beta^\mu + \Omega \phi^\mu$ is the corotating shift, and $t^\mu \equiv \alpha n^\mu + \beta^\mu$. 

The 4-velocity of the fluid is $u^\alpha=u^t(1,v^i)$ where $v^i$ is the Newtonian
velocity (velocity with respect to the inertial observers) that can be split into 
two parts: one that follows the rotation around the center of mass, 
$\Omega\phi^i$, and the velocity with respect to the corotating observer $V^i$, i.e. 
\begin{equation}
u^\alpha \equiv u^t (t^\alpha + v^\alpha) = u^t (k^\alpha + V^\alpha) .
\label{eq:4vel}
\end{equation}
where $V^\alpha=(0,V^i)$ and $v^\alpha=(0,v^i) \equiv\Omega\phi^\alpha+V^\alpha$.
For corotating BNSs, $V^i=0$.

\subsection*{The potential fluid equation}
\label{sec:pfe}

For an irrotational flow, the relativistic vorticity tensor 
\begin{equation}
\omega_{\alpha\beta}\equiv\nabla_{\alpha}(hu_\beta) - \nabla_{\beta}(h u_\alpha)\,   
\label{eq:rvt}
\end{equation}
is zero. Here $h$ is the relativistic specific enthalphy.
This implies that in such a case $h u_\alpha=\nabla_\alpha\Phi$ i.e. the 4-velocity can be
derived from a fluid potential $\Phi$. In the general case where the stars are arbitrarily
spinning, one can write
\begin{equation}
hu_\alpha=\nabla_\alpha \Phi + s_{\alpha} \,, 
\label{eq:hua_irr_spin}
\end{equation}
where $s_\alpha$ is their spin vector. In the decomposition above, we implicitly follow the discussion in~\cite{Tichy:2011gw,Tsokaros_2015} which is by no means unique. In a similar way, one can follow \cite{Tsokaros:2018dqs} where another decomposition based on the circulation has been proposed. In such case the potential equation has a different form although it still preserves the characteristics that motivate this work.

In 3+1 form, the components of interest of the velocity are the ones projected into the hypersurface
$\Sigma_t$ through the projection tensor $\gamma_{\mu\nu}=g_{\mu\nu}+n_\mu n_\nu$. Its pullback to 
$\Sigma_t$ is the tensor $\gamma_{ij}$. Defining
\begin{equation}
\hat{u}^i \equiv \gamma^i_\mu h u^\mu = h u^t(\omega^i + V^i),
\end{equation}
Eqs. (\ref{eq:hua_irr_spin}) and (\ref{eq:4vel}) imply 
\begin{equation}
    V^i = \frac{D^i\Phi +s^i}{h u^t} - \omega^i  =
    \frac{\delta^{ij}\partial_j\Phi / \psi^4 + s^i}{h u^t} - (\beta^i +\Omega\phi^i)  
\label{eq:corotvel}
\end{equation}
where $D_i$ is the covariant derivative associated with $\gamma_{ij}$, and $\delta^{ij}$ is the Kronecker delta.
Conservation of rest mass together with Eq. (\ref{eq:corotvel}) yields an elliptic equation for $\Phi$
\begin{eqnarray}
\partial_i \partial^i \Phi =& -\frac{2}{\psi}\partial_i \psi \partial^i\Phi + \psi^4 \omega^i\partial_i(h u^t)    
 - \psi^4\left(\partial_i s^i + s^i \partial_i\ln\psi^6\right)  \nonumber \\ 
&- \frac{h}{\alpha \rho}  (\partial^i\Phi + \psi^4 s^i - \psi^4 h u^t \omega^i)  \partial_i \frac{\alpha \rho}{h}   ,
\label{eqn:bns_Poisson}
\end{eqnarray}

where $\rho$ is the rest-mass density inside the star. We call Eq. (\ref{eqn:bns_Poisson}) the potential fluid 
equation. 
On the surface of the star we have $\rho=0$, therefore
evaluating Eq. (\ref{eqn:bns_Poisson}) there, we obtain a Neumann-type boundary condition \cite{Tsokaros_2015}
\begin{equation}
\left[ (\partial^i\Phi + \psi^4 s^i - \psi^4 h u^t \omega^i)\partial_i \rho \right]_{\rm surface} = 0  .
\label{eqn:bns_boundary}
\end{equation}
In the \textsc{cocal} solver, a numerical method to find the solution of the boundary problem above
is implemented and described in detail 
in~\cite{Tsokaros_2015, 2000PhRvD..61l4023U}. The method used in \textsc{cocal} inverts the Laplacian 
through the Green's function $G(x,x')=1/|x-x'|$ while adding a harmonic function such that the sum of 
the two will satisfy the boundary condition Eq.~(\ref{eqn:bns_boundary}). In addition, separate to the
normal coordinate system, fluid surface fitted coordinates are implemented. In section \ref{sec:result3} of this article, we present a comparison between the solution from the new approach using the source term method and the solution from \textsc{cocal}. 
Notice that the source term method does not need to implement fluid surface fitted coordinates.

It should be noted that in real BNS initial data, the variables in Eq.~(\ref{eqn:bns_Poisson}) and~(\ref{eqn:bns_boundary}) besides $\Phi$ (i.e. $h$, $u^t$, $\alpha$, etc.) are not known a-priori, and they also depend on $\Phi$. Specifically, five more elliptic equations~\cite{Tsokaros_2015} for $\psi$, $\alpha$, and $\beta^i$ would have to be solved simultaneously with Eq.~(\ref{eqn:bns_Poisson}). The elliptic equation for $\Phi$ is the primary focus in this article particularly because the boundary condition Eq.~(\ref{eqn:bns_boundary}) is imposed on the surface of the star. In contrast, the other five elliptic equations impose boundary conditions at spatial infinity where asymptotic flatness of spacetime is assumed~\cite{Tsokaros_2015}. For a complete BNS initial data solver, an iterative scheme for Eq.~(\ref{eqn:bns_Poisson}) as well as the other five equations is implemented such that each variable is solved independently while the others are fixed.

For simplicity, we will assume irrotational flows, i.e. $s^i=0$ in this work. 
All terms in the right-hand side of Eq.~(\ref{eqn:bns_Poisson}) are easily computed except the last one that includes a division by $\rho$,
which tends to zero at the surface of the NS. This ``problematic'' term exists even when $s^i=0$ and in the 
following sections we describe in detail how to treat it. The qualitative behavior of  Eq.~(\ref{eqn:bns_Poisson})
is not affected by the spin since in effect it only modifies the also regular shift terms (proportional to $\omega^i$)
both in Eq.~(\ref{eqn:bns_Poisson}) as well as in the boundary condition Eq.~(\ref{eqn:bns_boundary}).
Term $\partial_i s^i$ for NSs that are uniformly rotating is approximately constant, thus it does not influence the
solution method presented here. Therefore, including spin does not complicate the problem and can be trivially 
accounted for in the methods presented below.

\section{The source term method}
\label{sec:stm}
In this section, we will discuss a method by Towers~\cite{STM_2018}, called the source term method (STM), 
that solves the fluid potential boundary value problem (Eqs.~(\ref{eqn:bns_Poisson}),(\ref{eqn:bns_boundary}))
in a novel way that encapsulates the boundary condition on the surface of the star as jump conditions. Notably, by
embedding the domain of the star in a \textit{larger computational domain}, this method does not require a surface fitted 
coordinate system, thus providing an easy and robust alternative method to the BNS initial value problem.

The fluid potential BVP Eqs.~(\ref{eqn:bns_Poisson}),(\ref{eqn:bns_boundary}) can be written in general as
\begin{equation}
\cases{
    \rho > 0 \text{ (interior)}, & $\nabla^2 \Phi = S^+(\vec{x})$, \\
    \rho = 0 \text{ (surface)}, & $\Phi_n = a(\vec{x})$,
}
\label{eqn:pois}
\end{equation}
where
\begin{equation*}
\Phi_n \equiv \frac{\partial}{\partial n}\Phi \equiv \hat{n} \cdot \nabla \Phi 
\end{equation*}
is the derivative of $\Phi$ along the unit normal of the surface of the star $\hat{n}$, and $S^+(\vec{x})$ is the
right-hand side of  Eq.~(\ref{eqn:bns_Poisson}). 
Note that from this section onwards, we will denote by $\hat{n}$ the unit normal of the boundary (not to be confused
with the unit timelike normal $n^\mu$ introduced in Eq.~(\ref{eq:hkv2})) and  $\nabla$ the spatial derivative operator in Cartesian coordinates (not to be confused with the covariant derivative operator $\nabla_\mu$ in sec.~\ref{sec:phys_theory}).
In the real BNS initial value problem, the surface of the star is
not known a-priori and constitutes part of the solution. Here, we consider the problem where the density $\rho$ is given, which indicates
the surface where $\rho=0$ is known as well. The unit normal to the surface can be
written as $\hat{n}=-\nabla\rho/\abs{\nabla \rho}$, and thus the boundary condition Eq.~
(\ref{eqn:bns_boundary}) yields 
\begin{equation}
\Phi_n \equiv -\frac{\nabla\Phi \cdot \nabla\rho}{\abs{ \nabla\rho}} = 
-\frac{\psi^4 h u^t \vec{\omega} \cdot \nabla \rho}{\abs{ \nabla \rho}}
\label{eq:phin}
\end{equation}

The main idea of the source term method~\cite{STM_2018} is, instead of solving the Neumann BVP with an 
unknown boundary (the surface of the star), to embed these boundary conditions as jump conditions in
the source terms, $S^+$, and solve a generalized Poisson equation that contains extra
terms in addition to $S^+$ due to the boundary conditions on a \textit{larger computational grid}.
In effect, this procedure moves the unknowns from the boundary to the source terms of the Poisson
equations, and hence its name.
In the next section we describe how this procedure can be accomplished and the meaning of the 
larger computational domain. 

\begin{figure}
\centering
\includegraphics[width=7.5cm]{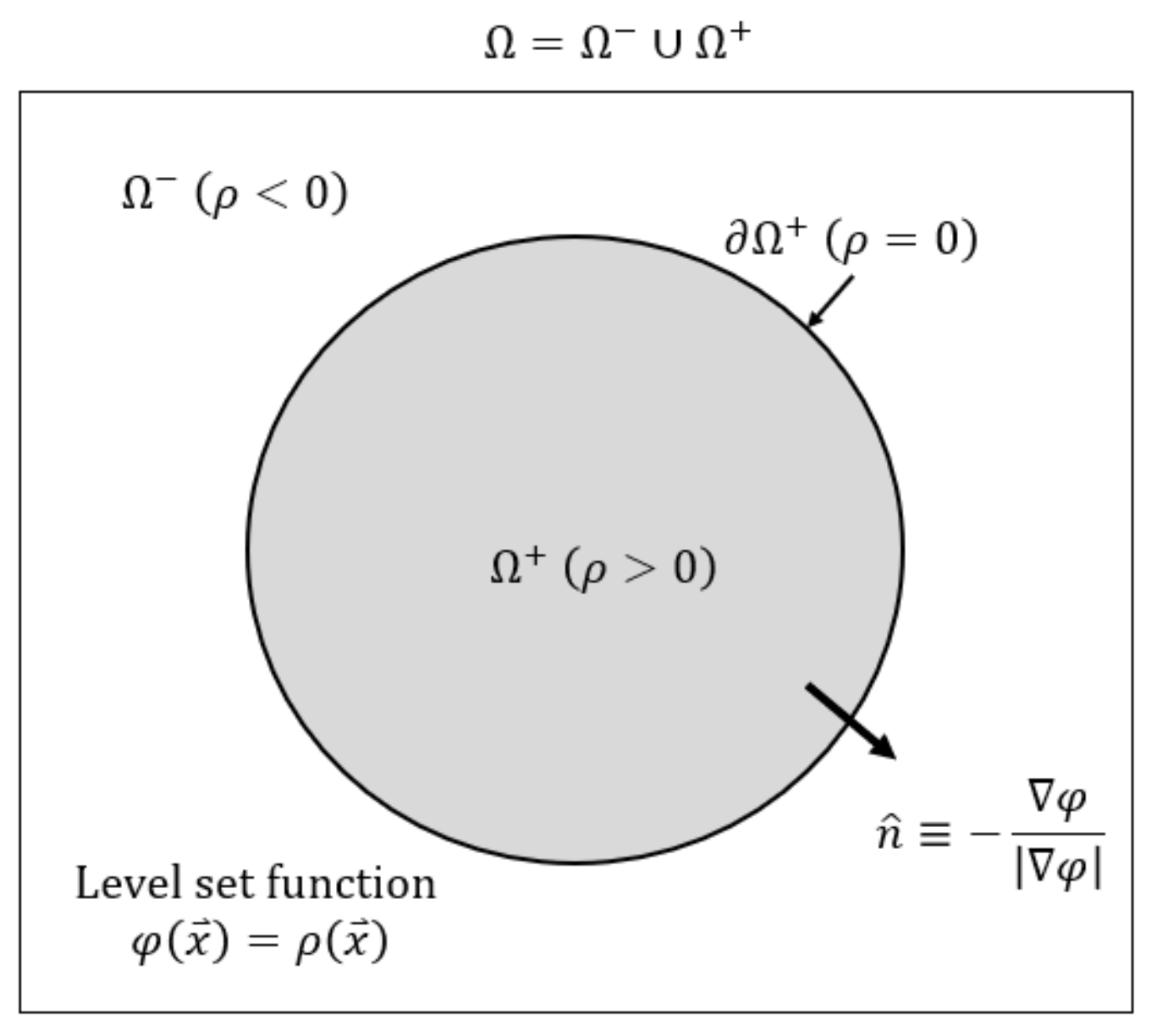}
\caption{The diagram shows the domain of interest $\Omega^+$ in grey with the boundary condition specified on $\partial \Omega^+$. Instead of solving the BVP on $\partial \Omega^+$, the boundary condition is embedded as jump conditions and the Poisson equation is solved over the larger domain $\Omega$.}
\label{fig:domain_1}
\end{figure}

\subsection{Domain embedding and the source term method equation}
\label{sec:domain_emb}

Consider the domain $\Omega = \Omega^+ \cup \Omega^-$ in Fig.~\ref{fig:domain_1} and the following BVP,
\begin{equation}
\cases{
\nabla^2 \Phi = S^+(\vec{x}), &  $\vec{x} \in \Omega^+$,  \\
\nabla^2 \Phi = S^-(\vec{x}), &  $\vec{x} \in \Omega^-$,  \\
\Phi = 0,                     &  $\vec{x} \in \partial\Omega$,    \\
[\Phi_n] = a(\vec{x}),        &  $\vec{x} \in \partial\Omega^+$,  \\
[\Phi] = b(\vec{x}),          &  $\vec{x} \in \partial\Omega^+$,  \\
}
\label{eq:gbvp}
\end{equation}
where we denote by $[\Phi] \equiv \Phi^+ - \Phi^- = b(\vec{x})$ and 
$[\Phi_n] \equiv \Phi_n^+ - \Phi_n^- = a(\vec{x})$. Here
$\Phi^+, \Phi_n^+$ are the solution $\Phi$ and its normal derivative $\Phi_n$ evaluated at the interior side 
of the boundary $\partial\Omega^+$, and $\Phi^-, \Phi_n^-$ are those evaluated at the
exterior side of $\partial\Omega^+$.
The interior of the star is represented by $\Omega^+$ and its boundary $\partial\Omega^+$
is defined through the level set function $\varphi (\vec{x})$ 
which typically can be taken to be the rest-mass density $\rho(\vec{x})$ of the star 
(as in Fig.~\ref{fig:domain_1}) and is used to characterize the various regions of interest.
\begin{equation}
\cases{
\varphi > 0, & $\vec{x} \in \Omega^+$, \\
\varphi < 0, & $\vec{x} \in \Omega^-$, \\
\varphi = 0, & $\vec{x} \in \partial{\Omega^+}$.
}
\label{eqn:Poisson}
\end{equation}
Notice that although $\rho<0$ does not exist in a BNS problem, we can still make such an 
assumption in order to establish the level set function Eq. (\ref{eqn:Poisson}).
The unit normal vector corresponding to the level set function is thus
\begin{equation}
\hat{n} \equiv -\frac{\nabla \varphi}{\abs{\nabla\varphi}}.
\label{eq:normalphi}
\end{equation}

The source term method treats the boundary condition at $\partial\Omega^+$ 
as a source term and solves the equation in the 
larger rectangular grid $\Omega = \Omega^+ \cup \Omega^-$. This allows the boundary conditions on 
$\partial \Omega^+$ to become jump conditions that can be absorbed into the sources. The generalized Poisson equation (see Appendix~\ref{appdx:stm_derivation} for a derivation)
that will be solved in the whole domain $\Omega$ becomes
\begin{equation}
\label{eqn:source_term}
\nabla^2 \Phi = \nabla^2(bH) - H\nabla^2 b - \left(a -\frac{\partial b}{\partial n}\right)\abs{\nabla\varphi}\delta(\varphi) + S,
\end{equation}
where $S(\vec{x})=S^+ H(\varphi(\vec{x})) + S^-(1-H(\varphi(\vec{x})))$, $H(\varphi)$ is the
Heaviside step function, and $\delta(\varphi)$ is the Dirac delta function.

In the context of the fluid potential Eq.~(\ref{eqn:pois}), the terms $S^-$ and $\Phi^-$ are freely specifiable and we choose $S^-=0$ and $\Phi^-=0$ to simplify 
the jump conditions $a(\vec{x})$, $b(\vec{x})$. To summarize, the coefficients in Eq.~(\ref{eqn:source_term}) are:
\begin{equation}
    \label{eq:five_var}
    \fl \cases{
        S^+(\vec{x}) &$= -\left(\frac{2}{\psi}\nabla \psi + \nabla \left(\ln{\frac{\alpha \rho}{h}} \right)\right) \cdot \nabla \Phi^+
        + \psi^4 \vec{\omega}\cdot\nabla(h u^t) + \psi^4 h u^t \vec{\omega} \cdot \nabla \left(\ln{\frac{\alpha \rho}{h}}\right)$, \\
        S^-(\vec{x}) &= $0$, \\
        a(\vec{x}) &$= \Phi_n^+ - \Phi_n^- = \Phi_n^+ = -\frac{\psi^4 h u^t \vec{\omega}\cdot \nabla \rho}{\abs{\nabla \rho}}$, \\
        b(\vec{x}) &$= \Phi^+ - \Phi^- = \Phi^+$, \\
        \varphi(\vec{x}) &$= \rho$.
    }
\end{equation}

For a Neumann BVP, the values for $b(\vec{x})$ and $S^+(\vec{x})$ remain unknown as they depend on $\Phi$. This problem is addressed in~\cite{STM_2018}, and an iterative scheme is proposed. The iterative scheme starts with an initial $\Phi = 0$. During each iteration, the source term method solves for the Poisson equation with Eq.~(\ref{eqn:source_term}), and the coefficients in Eq.~(\ref{eq:five_var}) are then updated based on the current $\Phi$ from the solver. The process continues until the solution $\Phi$ converges.

In order to simplify our code, Eq.~(\ref{eqn:source_term}) can be rewritten as a variable coefficient Poisson equation with the same boundary condition Eq.~(\ref{eq:phin}):
\begin{equation}
\cases{
\nabla \left(\zeta(\vec{x}) \nabla \Phi \right) = S^+_{\zeta}(\vec{x}), &$\vec{x} \in \Omega^+$, \\
\nabla \left(\zeta(\vec{x}) \nabla \Phi \right) = S^-_{\zeta}(\vec{x}), &$\vec{x} \in \Omega^-$, \\
\Phi_n = a(\vec{x}), & $\vec{x} \in \partial \Omega^+$,
}
\label{eqn:coef_Poisson}
\end{equation}
where
\begin{equation*}
\cases{
\zeta(\vec{x}) & $= \frac{\psi^2\alpha\rho}{h}$, \\
S^+_{\zeta}(\vec{x}) & $=\psi^6\vec{\omega}\cdot\nabla(\alpha\rho u^t)$, \\ 
S^-_{\zeta}(\vec{x}) & $= 0$. \\ 
}
\end{equation*}
The source terms are being related through 
$S^+ = (S^+_\zeta - \nabla \zeta \cdot \nabla \Phi)/\zeta$.
This formulation is used for two reasons: a) $S^+_\zeta$ does not depend on $\Phi$, thus the 
iterative source term method can be simplified by taking only $\zeta, S^+_\zeta, a, \varphi$ as 
input variables; b) in the Newtonian limit, $\zeta(\vec{x}) = \rho(\vec{x})$, so the algorithm 
is further simplified by taking $\zeta, S^+_\zeta, a$ as the only input variables.

\section{Numerical discretization}
\label{sec:numerical_discretization}

In this work, we follow the algorithm and numerical discretization for the source term method as described in~\cite{STM_2018}.
For simplicity, we use the notation $Z_{i,j,k} \equiv Z(x_i,y_j,z_k)$ for any field $Z$ defined on the grid.
The grid has spacing  $\Delta x = \Delta y = \Delta z$ with $(N_{grid} + 1)$ points in every direction.
We use second order finite differences for the partial derivatives and the Laplacian in 3D Cartesian coordinates. 

\begin{eqnarray*}
\fl \frac{\partial Z_{i,j,k}}{\partial x} = \frac{1}{2\Delta x}(Z_{i+1,j,k}-Z_{i-1,j,k}), \\
\fl\nabla^2 Z_{i,j,k} = \frac{1}{\Delta x^2}(Z_{i+1,j,k}+Z_{i-1,j,k}+Z_{i,j+1,k} 
 + Z_{i,j-1,k}+Z_{i,j,k+1}+Z_{i,j,k-1}-6Z_{i,j,k}). \\
\end{eqnarray*}
            
\begin{figure}
\centering
\includegraphics[width = \textwidth]{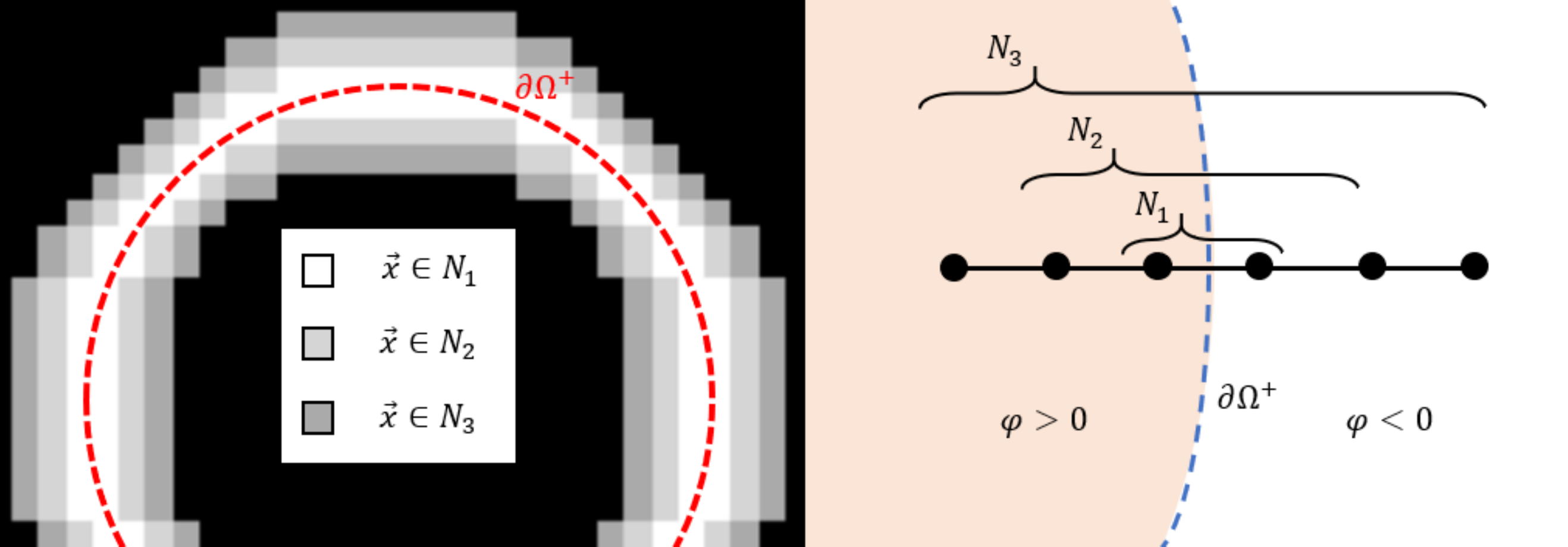}
\caption{Diagrams showing the neighboring points for the discretization of the source term method. 
The image on the left is a two-dimensional diagram showing $N_1$, $N_2$, $N_3$ based on the boundary 
$\partial \Omega^+$. The image on the right demonstrates a slice of the two-dimensional plot where 
$N_1 \subset N_2 \subset N_3$.}
\label{fig:neighbor}
\end{figure}

For the source term in Eq.~(\ref{eqn:source_term}), a discretization of the Heaviside and delta function 
is required. Here we use the discretization provided in~\cite{Heaviside_2009} and repeat the salient points for completeness of presentation.
Since the boundary $\partial \Omega^+$ is defined through the level function $\varphi$, 
we first define the set $N_1$ that contains the grid points on or adjacent to $\partial \Omega^+$ 
(see Fig.~\ref{fig:neighbor}) as 
\begin{eqnarray*}
    N_1 = \{(x_i,y_j,z_k)\ |\ &\varphi(x_{i\pm1},y_j,z_k)\varphi(x_i,y_j,z_k) \leq 0\ \text{or} \\ 
    &\varphi(x_i,y_{j\pm1},z_k)\varphi(x_i,y_j,z_k) \leq 0\ \text{or} \\ 
    &\varphi(x_i,y_j,z_{k\pm1})\varphi(x_i,y_j,z_k) \leq 0\ \} .
\end{eqnarray*}

The characteristic function $\chi(\vec{x})$ of the set $N_1$ is defined to be
\begin{equation}
\chi(\vec{x}) = 
\cases{
1 , & $\vec{x} \in N_1$, \\
0 , & otherwise. \\}
\end{equation}
Further we define the neighboring sets $N_2, N_3,\ldots$ as
\begin{equation}
N_2 = \Xi(N_1), \quad N_3 = \Xi(N_2), \quad \ldots
\end{equation}
where $\Xi(P)$ is a function that returns the set of grid points on or adjacent to the set of points $P$.
\begin{equation*}
\Xi(P) \equiv P \cup \{(x_i,y_j,z_k)\ |\ (x_{i\pm1},y_j,z_k)
        \text{ or } (x_i,y_{j\pm1},z_k) \text{ or } (x_i,y_j,z_{k\pm1}) \in P \}.
\end{equation*}
The neighboring sets $N_1, N_2, N_3, ...$ are visualized in Fig.~\ref{fig:neighbor}, and these sets are used in this section and beyond to specify the domain used in the implementation of the source term method.
        
For simplicity of the discretization formulas, we define functions 
$I(\varphi(\vec{x}))$, $J(\varphi(\vec{x}))$, and $K(\varphi(\vec{x}))$ as
\begin{equation}
\label{eq:IJK}
\fl I(\varphi(\vec{x})) = {\rm max}(0,\varphi(\vec{x})), \quad
J(\varphi(\vec{x})) = \frac{1}{2}I(\varphi(\vec{x}))^2, \quad
K(\varphi(\vec{x})) = \frac{1}{6}I(\varphi(\vec{x}))^3, 
\end{equation}
whose gradients can be written as 
$\nabla I(\varphi) = H(\varphi)\nabla\varphi$, 
$\nabla J(\varphi) = I(\varphi)\nabla\varphi$, and 
$\nabla K(\varphi) = J(\varphi)\nabla\varphi$.
The discretization of the Heaviside function $H^h_{i,j,k}$ for $H(\varphi)$ and the delta function 
$\delta^h_{i,j,k}$ for $\delta(\varphi)$ are then expressed~\cite{STM_2010} as (see Appendix~\ref{appdx:heaviside_delta_derivation} for a derivation)
\begin{eqnarray}
H^h_{i,j,k}\ =\ & \chi_{i,j,k} \left( \frac{\nabla^2 J_{i,j,k}}{\abs{\nabla \varphi_{i,j,k}}^2} - 
\frac{(\nabla^2 K_{i,j,k} - J_{i,j,k} \nabla^2 \varphi_{i,j,k})\nabla^2 \varphi_{i,j,k}}{\abs{\nabla \varphi_{i,j,k}}^4} \right) \nonumber \\
& +\   (1 - \chi_{i,j,k}) H(\varphi_{i,j,k})  \label{eq:Heaviside},  \\
\qquad \nonumber\\
\delta^h_{i,j,k}\ =\ & \chi_{i,j,k} \left( \frac{\nabla^2 I_{i,j,k}}{\abs{\nabla \varphi_{i,j,k}}^2} - 
\frac{(\nabla^2 J_{i,j,k} - I_{i,j,k} \nabla^2 \varphi_{i,j,k})\nabla^2 \varphi_{i,j,k}}{\abs{\nabla \varphi_{i,j,k}}^4} \right).   \label{eq:Delta}
\end{eqnarray}

Given the discretization for $H^h_{i,j,k}$ Eq.~(\ref{eq:Heaviside}) and $\delta^h_{i,j,k}$ 
Eq.~(\ref{eq:Delta}), one can proceed with the disctretization of Eq.~(\ref{eqn:source_term}) (note that $H^h_{i,j,k}$ is different from $H_{i,j,k} = H(\varphi_{i,j,k})$):
\begin{eqnarray}
\fl \nabla^2 \Phi_{i,j,k} =\ & \nabla^2(b_{i,j,k}H(\varphi_{i,j,k})) - H^h_{i,j,k}\nabla^2 b_{i,j,k} - \left(a_{i,j,k} -
\frac{\partial b_{i,j,k}}{\partial n}\right)|\nabla\varphi_{i,j,k}|\delta^h_{i,j,k} \nonumber \\
 & +\  H^h_{i,j,k}S^+_{i,j,k}.
\label{eq:stm_discretized}
\end{eqnarray}
                
\subsection{Boundary treatment}
\label{sec:boundary_extension}

When solving for the Poisson equation using Eq.~(\ref{eqn:source_term}), the boundary condition 
$a(\vec{x}) = \Phi_n^+(\vec{x})$ (Eq.~(\ref{eq:phin})) needs to be imposed. There are two difficulties
here. First, the boundary does not coincide with grid points in a typical case. Second, in the realistic
case of BNSs, the boundary is not known a priori, but only when a solution is found.
In other words, the boundary is changing through the iteration process. Both these problems imply that
it is not trivial to apply boundary condition Eq.~(\ref{eq:phin}) for the Poisson equation. In order to address these problems, we compute the so-called ``projected points'' $\{x^p,y^p,z^p\}$ \cite{STM_2018}
which in effect represent the 
numerical boundary where the boundary condition Eq.~(\ref{eq:phin}) is evaluated. These locations
do not belong to the grid and are close to the true boundary $\partial \Omega^+$.  
\begin{figure*}[!htbp]
\centering
\begin{minipage}{.45\textwidth}
\centering
\includegraphics[width = \linewidth]{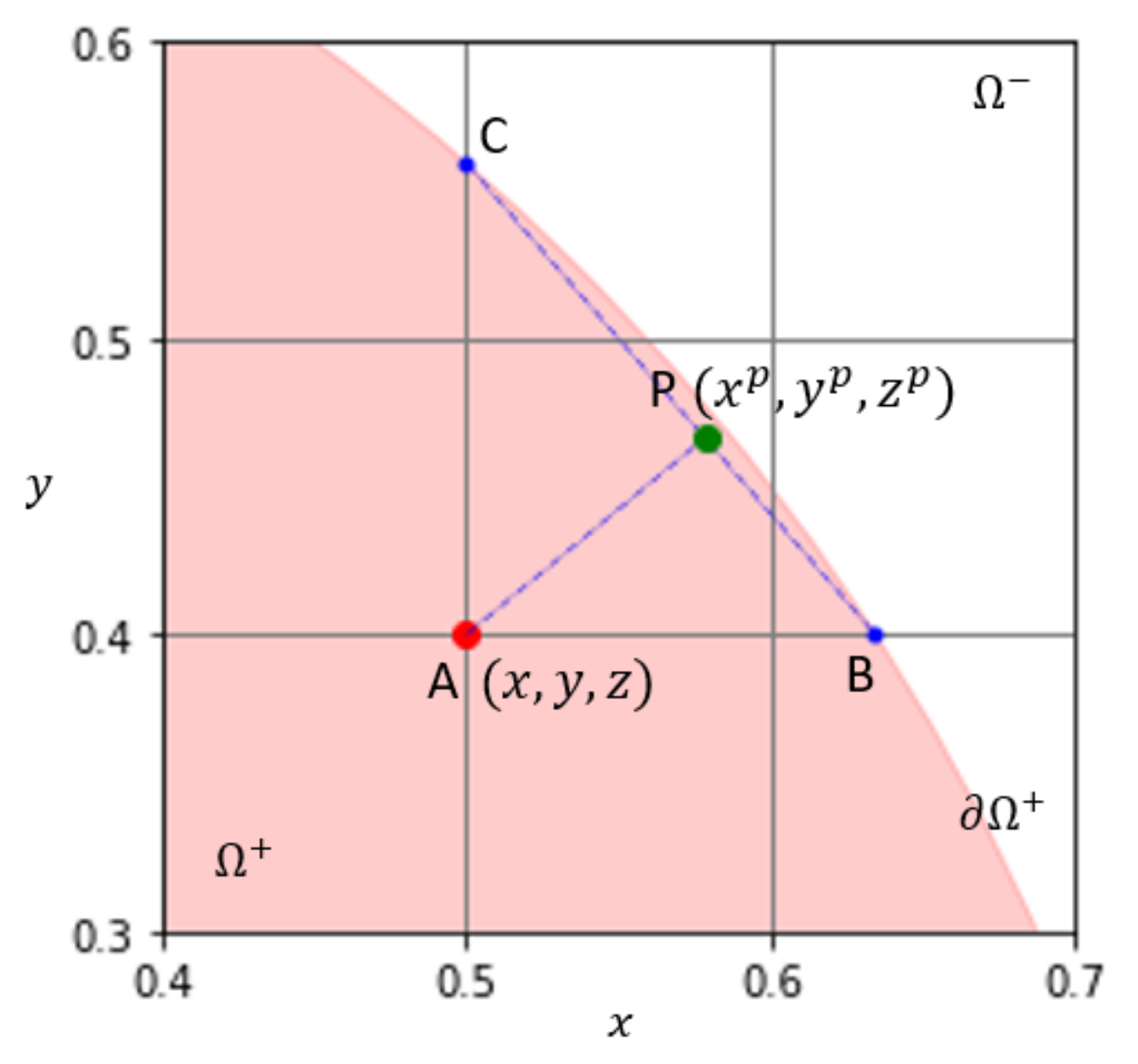}
\caption{A diagram of the boundary extension of the grid points near the boundary. Assuming the level 
set $\varphi$ is decreasing trilinearly near the boundary, the projected point $P$ 
is the projection of the original grid point $A$ onto the plane determined by the projected points of $A$ on the boundary $\partial \Omega^+$ in the 
$x,y,z$ direction (as indicated by point $B$ and $C$ in the 2D diagram).}
\label{fig:boundaryExtension}
\end{minipage}
\qquad
\begin{minipage}{.45\textwidth}
\centering
\includegraphics[width=\linewidth]{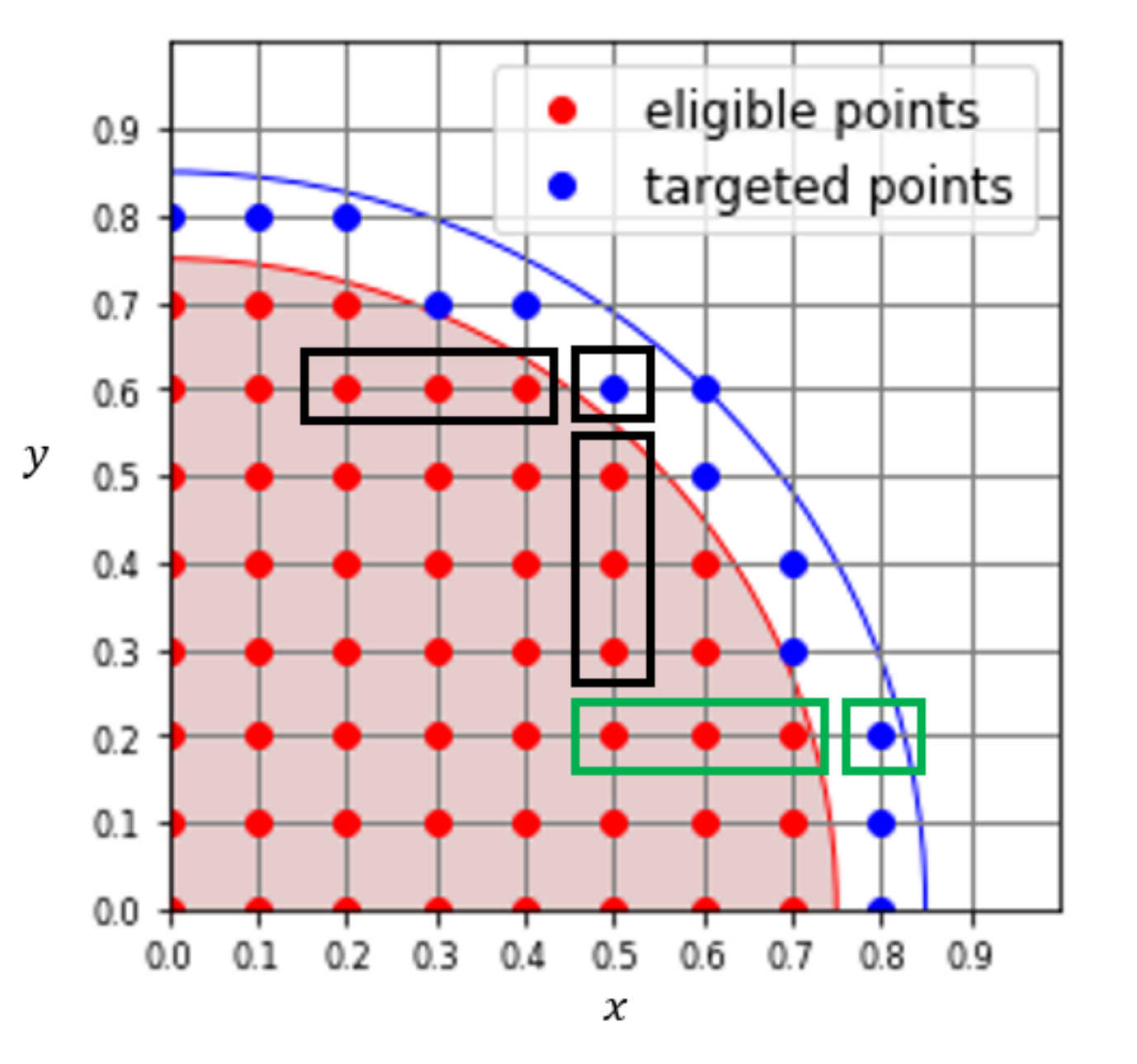}
\caption{A diagram of the extrapolation method. For the target point at $(0.8, 0.2)$, the returned value is 
extrapolated from the three stencil points left to itself; for the target point at $(0.5, 0.6)$, the returned 
value is the average of the extrapolated value from the three stencil points to its left and the extrapolated 
value from the three stencil points underneath.}
\label{fig:extpl}
\end{minipage}
\end{figure*}

Using a first order Taylor approximation, the projected points can be written as 
\begin{eqnarray*}
x^p = x + \frac{\varphi}{\abs{\nabla\varphi}}\hat{n}_x, 
\quad y^p =& y + \frac{ \varphi}{\abs{\nabla\varphi}}\hat{n}_y,
\quad z^p = z + \frac{ \varphi}{\abs{\nabla\varphi}}\hat{n}_z, \\
\text{where} \ \hat{n}_x = \frac{-\partial_x \varphi}{\abs{\nabla\varphi}} ,\quad \hat{n}_y =& \frac{-\partial_y \varphi}{\abs{\nabla\varphi}} ,\quad \hat{n}_z = \frac{-\partial_z \varphi}{\abs{\nabla\varphi}} \ .
\end{eqnarray*}
Fig.~\ref{fig:boundaryExtension} shows how the projected points are calculated from the grid points.

The extended boundary condition at each grid points $a^p(\vec{x})$ that will be used as the boundary condition in our code takes the value of $a(\vec{x})$ evaluated at the projected points $a^p(x,y,z) = a(x^p,y^p,z^p)$.
For test cases where the boundary 
function $a(\vec{x})$ is given, the extended boundary condition $a^p$ can be directly calculated; for BNSs where the boundary 
condition is given as an array of values, an interpolation must be performed first to obtain the boundary 
function. A more detailed description of the effect of the extended boundary treatment on various types of BVP can be found in \cite{STM_2018}.
        
\subsection{Extrapolation method}
\label{sec:extrapolation}

For the finite difference discretization of the source term method, the values of $\Phi$ and $\Phi_n$ are 
required to be defined throughout $N_2$ to avoid discrete jumps in the derivatives. Thus, an extrapolation 
method is needed to get the values outside the boundary in $N_2 / \Omega^+$. Specifically, a quadratic 
extrapolation method used in~\cite{STM_2018} is implemented, as shown in Fig.~\ref{fig:extpl}. 
        
The quadratic extrapolation algorithm is implemented for any given field $v(\vec{x})$ to find its extrapolated 
value $v_{ext}$, where in the source term method $v$ can be $\Phi$ or $\Phi_n$. The algorithm takes in three inputs: 
a) an initial array of values $v_0$; b) targeted points $\mathcal{T}_0$ that represent the initial set of 
points where the values are not defined and need to be extrapolated; c) eligible points $\mathcal{E}_0$ 
that represent the initial set of grid points where the values are defined and available to use for 
extrapolation. The extrapolation algorithm uses an iterative method to extrapolate the points adjacent to the eligible 
points layer by layer until all targeted points have an extrapolated value. First, the algorithm initializes the variables 
$\mathcal{T} = \mathcal{T}_0$, $\mathcal{E} = \mathcal{E}_0$ and, $v = v_0$. In each iteration, for a target 
point $\vec{x} \in \mathcal{T}$, the algorithm checks, in the $\pm x, \pm y, \pm z$ direction, if all three stencil points are eligible points (as shown in Fig.~\ref{fig:extpl}), also expressed as an eligible direction $d$, and the 
extrapolated value $v_{ext}^{(d)}$ can be evaluated with quadratic extrapolation. For example, let $x_1$, $x_2$, $x_3$ be the three stencil points in the $+x$ direction, the extrapolated value $v_{ext}$ is given with quadratic extrapolation
\begin{equation}
v_{ext} = 3 v(x_1) - 3v(x_2) + v(x_3).
\end{equation}
To be specific, for a grid point $(x_i, y_j, z_k)$, the three stencil points in the $+x$ direction are $x_1=(x_{i+1},y_j,z_k)$, $x_2 = (x_{i+2},y_j,z_k)$, $x_3 = (x_{i+3},y_j,z_k)$. Similarly, in the $-x$ direction, the stencil points are $x_1=(x_{i-1},y_j,z_k)$, $x_2 = (x_{i-2},y_j,z_k)$, $x_3 = (x_{i-3},y_j,z_k)$.
        
If there is at least one eligible direction, then the grid point is eliminate from target points $\mathcal{T}$ and added to eligible points $\mathcal{E}$, and the value of the grid point $v$ is updated to the average of all the extrapolated values from all eligible directions. The process continues until the set of target points becomes empty. 

\subsection{Coefficient singularity on the surface}
\label{sec:Num_coef_singular}

\begin{figure}[!htbp]
\centering
\includegraphics[width=7cm]{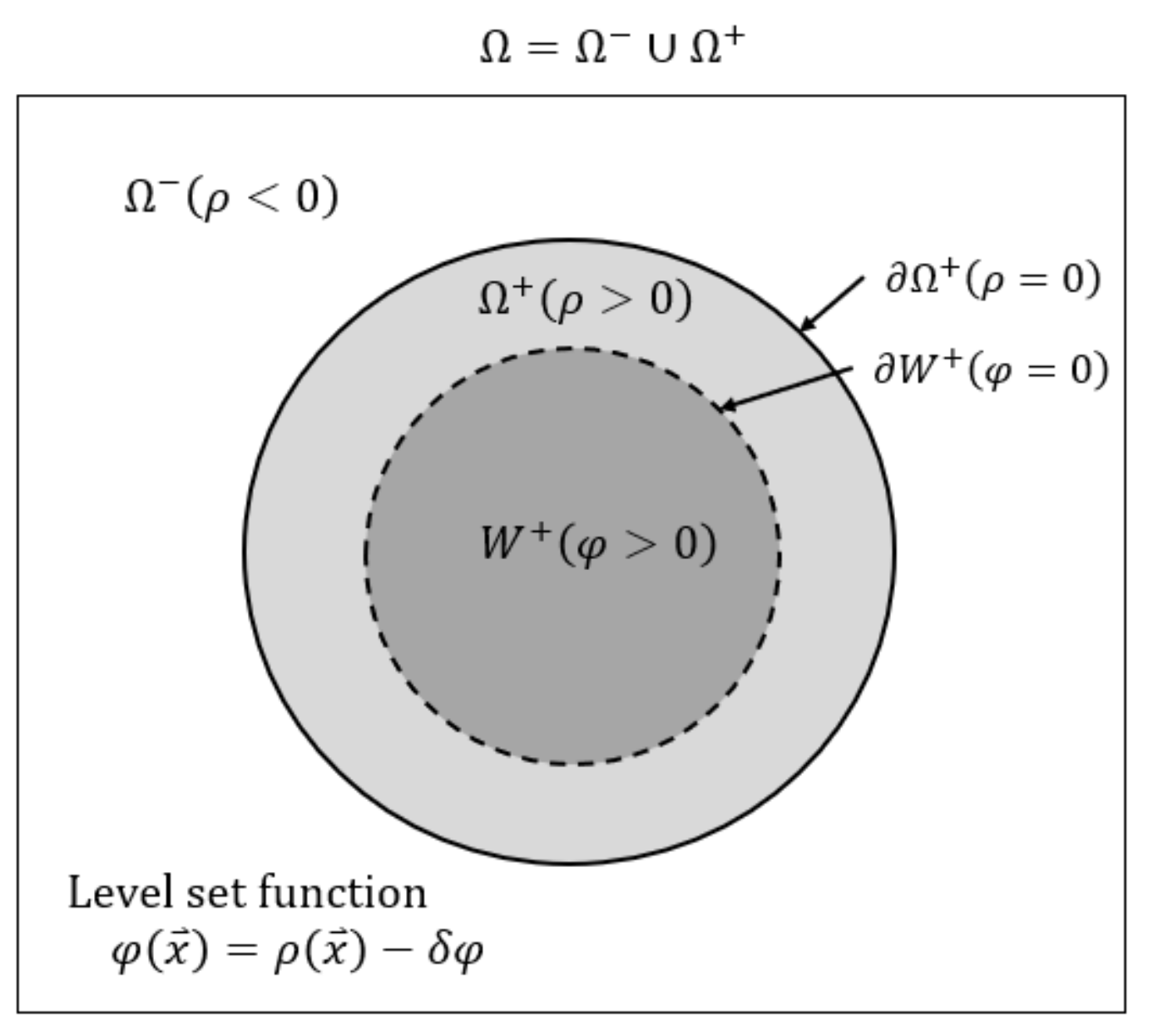}
\caption{The diagram of the interior $\Omega^+$ and the domain $W^+$ that the solver actually solves for. 
An additional separation $\delta\varphi$ is added to the level set function to shift the boundary inwards 
in order to avoid the singularity of the coefficient $\zeta$.}
\label{fig:domain-2}
\end{figure}

It is an intuitive choice to have the density $\rho$ to be the level set function since 
the surface is defined where $\rho = 0$. However, this raises a singularity issue in the coefficient 
$\zeta$. As $\rho \rightarrow 0$ near the surface, $\zeta = \psi^2\alpha\rho/h \rightarrow 0$. This 
becomes troublesome as the second order derivative in the equation vanishes, and a dividing by zero
occurs when evaluating $S^+ = (S^+_\zeta - \nabla \zeta \cdot \nabla \Phi) / \zeta$.
Our solution is to shift the level set with a constant separation $\delta \varphi$:
\begin{equation}
\varphi(\vec{x}) = \rho(\vec{x}) - \delta\varphi  .
\label{eq:levelsf}
\end{equation}
The additional separation introduces a domain $W^+\equiv\{\vec{x} | \varphi(\vec{x})>0\}$ that is slightly 
smaller than the original domain $\Omega^+\equiv\{\vec{x} | \rho(\vec{x})>0\}$ (as shown in Fig.~\ref{fig:domain-2}).

The iterative source term method will solve for $\Phi$ in the smaller domain $W^+$, and then the solution 
is extrapolated from $W^+$ to the entire domain $\Omega^+$ to get the final solution $\Phi_{ext}$. Nevertheless, 
this additional separation in the domain also changes the boundary condition $a(\vec{x})$ from $\partial \Omega^+$ to $\partial W^+$, which will introduce increasing error as $\delta\varphi$ increases. Thus, the value of 
$\delta \varphi$ should be carefully chosen such that it is not so small that the singularity in $\Omega^+$ causes the 
iterative scheme to diverge, and it is not so large that the change in boundary condition $a(\vec{x})$ is no 
longer consistent. Typically, we choose the separation $\delta\varphi$ to be a portion of the central density $\rho_0$, such as $\delta\varphi = 0.05\rho_0$. The choice for the value of the separation will be discussed in more detail in sec.~\ref{sec:result2}.
        
A detailed description of the pseudocode for the source term method algorithm is presented in~\cite{STM_2018}. For the numerical experiments in the next section, the iterative source term method terminates when the maximum relative change of the solution between successive iterations is less than $0.01\%$. For each iteration that does not terminate, the new solution is updated with a relaxation ratio of $0.1$. For the fast Poisson solver for the entire domain $\Omega = \Omega^+ \cup \Omega^-$, we use the Jacobi iterative method that terminates when the relative change of the method in the $2$-norm of the solution between successive iterations is less than $0.001\%$.

\section{Numerical results}
\label{sec:nr}

In this section, numerical experiments are conducted to test the performance of the source term method. 
We use the following rule for calculating the $L_2$ error of the result defined in the domain $\Omega^+$ :
\begin{equation}
\label{eqn:L2error}
E_{2} \equiv \sqrt{\frac{1}{N_{\Omega^+}}\sum_{\vec{x} \in \Omega^+}{(\Phi - \Phi_{\text{theory}})^2}},
\end{equation}
where $\Phi$ is the result of the solver, $\Phi_{\text{theory}}$ is the theoretical solution, and 
$N_{\Omega^+}$ is the number of grid points in the domain $\Omega^+$. It should be noted that since $\Phi$ 
is a solution of a Neumann BVP, the result is calibrated by subtracting the value at the center grid 
point to ensure a unique solution.
    
Section~\ref{sec:result1} presents a test case in 2D with a regular coefficient, while section~\ref{sec:result2} 
shows 2D test cases with singular coefficients. Table~\ref{tbl:test_case} shows our choices
for the velocity potential $\Phi$, the
coefficient function $\zeta$, and the density $\rho$, which defines the level set function $\varphi$. The choice of the coefficient $\zeta$ was motivated by the need to probe
different singularity behaviors as $\sqrt{r}$ and $r$. For synthetic cases in sec.~\ref{sec:result1} and~\ref{sec:result2}, the resolution $N_{grid}$ is varied in order to observe the convergence rate of the source term method, and the set of values for the resolution are selected to be even numbers ranging 
from $32\times32$ to $246\times246$. The grid is then set up with $(x,y) \in [-1,1]\times[-1,1]$, which has spacing $\Delta x = \Delta y = 2/N_{grid}$. 
Finally, in section~\ref{sec:result3}, the solution of the BVP 
Eqs.~(\ref{eqn:bns_Poisson}),(\ref{eqn:bns_boundary}) of a realistic 3D BNS is computed 
and compared with the corresponding solution of the \textsc{cocal} code.

\begin{table}[!htbp]
\centering
\begin{tabular}{|p{0.1\linewidth} | p{0.2\linewidth} | p{0.15\linewidth} | p{0.15\linewidth} |}
\hline 
No. &       $\Phi(x,y)$        & $\rho(x,y) $        &  $\zeta(x,y)$  \\ \hline
$1$   & $1+ \sin(x)\cos(y)$ & $\bar{r}^2 - r_0^2$  & $e^{r^2}$    \\ 
$2.1$   & $ y + \sin(x)$      & $r - r_0$      & $\rho$       \\ 
$2.2$   & $ y + \sin(x)$      & $\sqrt{r}- \sqrt{r_0}$      & $\rho$       \\ \hline
\end{tabular}
\caption{Functions and parameters for 2D test cases. For test case~$1$, the modified radius 
$\bar{r}\equiv \sqrt{(\epsilon x)^2+y^2}$ is used to form an elliptic domain with eccentricity 
$\epsilon = 1.2$, and $r_0 = \sqrt{e/4} \approx 0.824$. Test cases $2.1$ and $2.2$ have
singular coefficient near the boundary ($\zeta, \rho \rightarrow 0$).}
\label{tbl:test_case}
\end{table}

\subsection{Two-dimensional test with non-singular coefficient}
\label{sec:result1}
\begin{figure}
    \centering
    \includegraphics{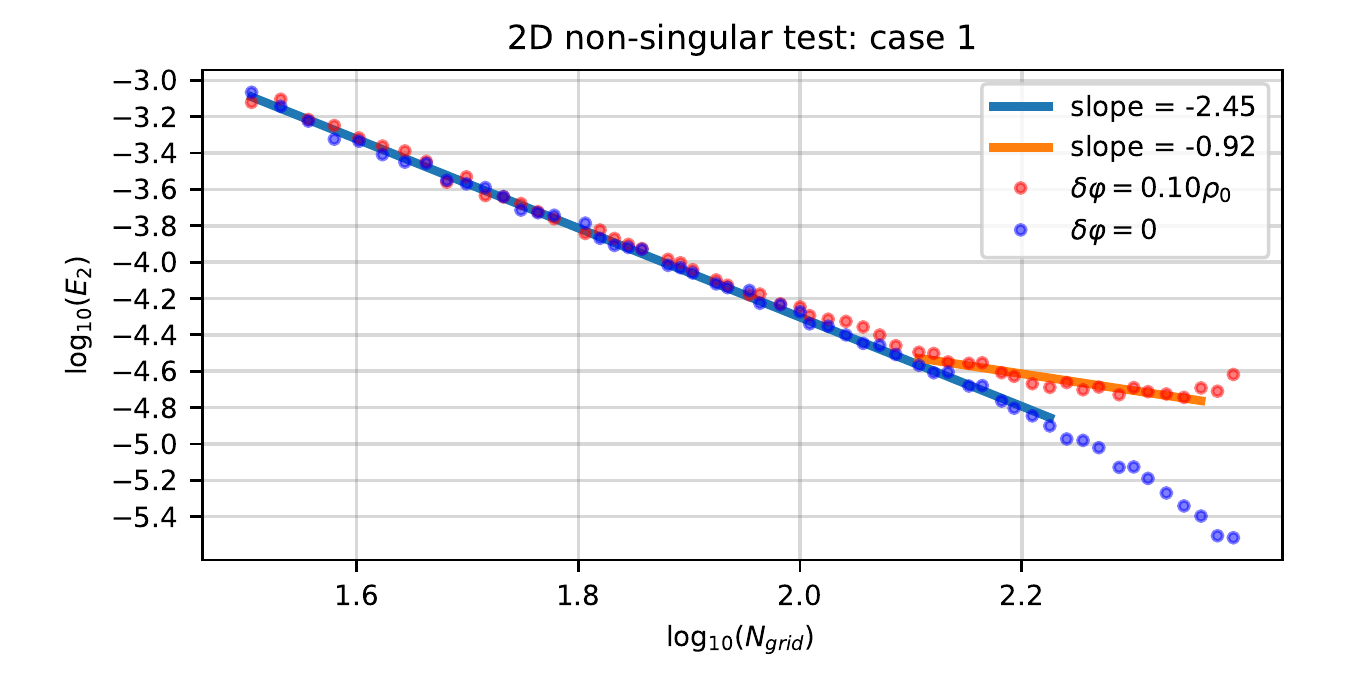}
    \caption{Convergence plot for 2D non-singular test case (test case~$1$). The plots shows the result: a) with no separation (data points in blue), which has a convergence rate of $2.45$; b) with a separation $\delta \varphi = 0.10 \rho_0$ (data points in orange).}
    \label{fig:plot_1-1}
\end{figure}

The result of the two-dimensional test case with non-singular coefficient (case~$1$ in Table~\ref{tbl:test_case}) shows an overall convergence of order $\sim 2.45$ 
as shown in Fig.~\ref{fig:plot_1-1}. Since in this example we have a non-singular coefficient, 
the treatment of additional separation is not required here. Notwithstanding that and in order 
to assess its effect in convergence we made a similar experiment with separation 
$\delta\varphi = 0.10\rho_0$. Fig.~\ref{fig:plot_1-1} shows the difference in convergence
between the solver with separation $\delta\varphi = 0.10 \rho_0$ and without any separation 
$\delta\varphi = 0$. The effect of a finite separation becomes apparent when the $N_{grid}=128$
where the rate drops to $\sim 0.92$ until $N_{grid}=238$. 
Beyond $N_{grid}=238$ or $\Delta x\sim 0.0084$ convergence is lost. As described in sec.~\ref{sec:Num_coef_singular}, the separation introduces an error in the boundary condition $a(\vec{x})$, thus resulting in the drop of the convergence rate.
    
\subsection{Two dimensional tests with singular coefficients}
\label{sec:result2}

As mentioned in sec.~\ref{sec:Num_coef_singular}, zeroes in the coefficient $\zeta$ near the boundary $\partial \Omega^+$ result in an instability. Therefore, an additional 
separation in the level set $\varphi = \rho - \delta \varphi$ is proposed at the cost of 
accuracy in the final result. In this section, such cases with singular coefficients 
are tested with various resolutions.
For simpler visualization of the separation, two test cases $2.1$ and $2.2$ in Table \ref{tbl:test_case} were selected with circularly symmetric coefficients $\zeta(r) = \rho(r) = r_0^\kappa - r^\kappa$, where $\kappa = 0.5,1.0$. Since both $\rho$ and $\varphi$ are circularly symmetric, we can define $\delta r$ to be the spatial separation between $\partial \Omega^+$ and $\partial W^+$ in any direction. Given the value for $\delta r$, one can find the separation $\delta \phi = (r_0^\kappa - (r_0 - \delta r)^\kappa)$ for test case~$2.1$ and $2.2$. Additionally, we also define the separation number $N_{sep}$ as the number of grid points across the separation:
\begin{equation}
    N_{sep} \equiv \frac{\delta r}{\Delta x}
\end{equation}

\begin{figure}[!htbp]
    \centering
    \includegraphics{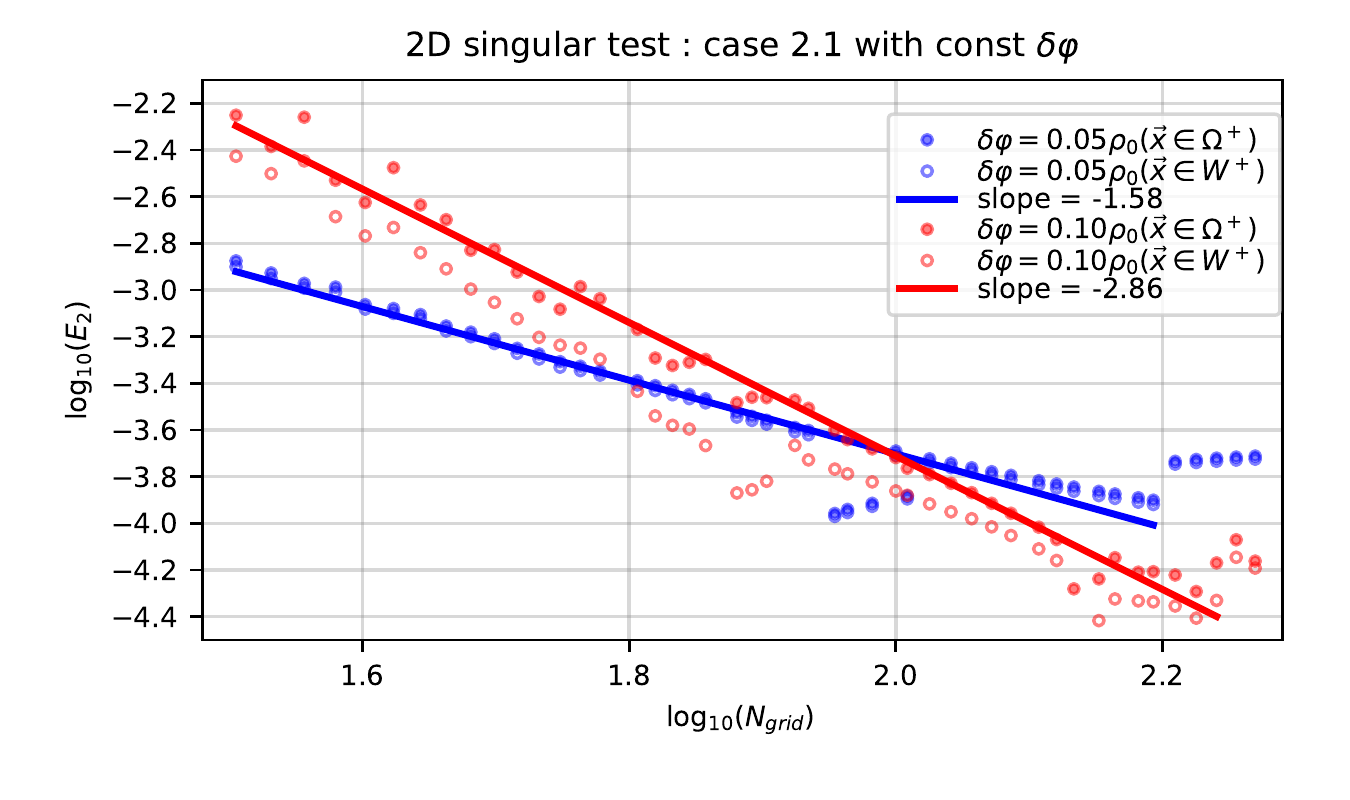}\\
    \includegraphics{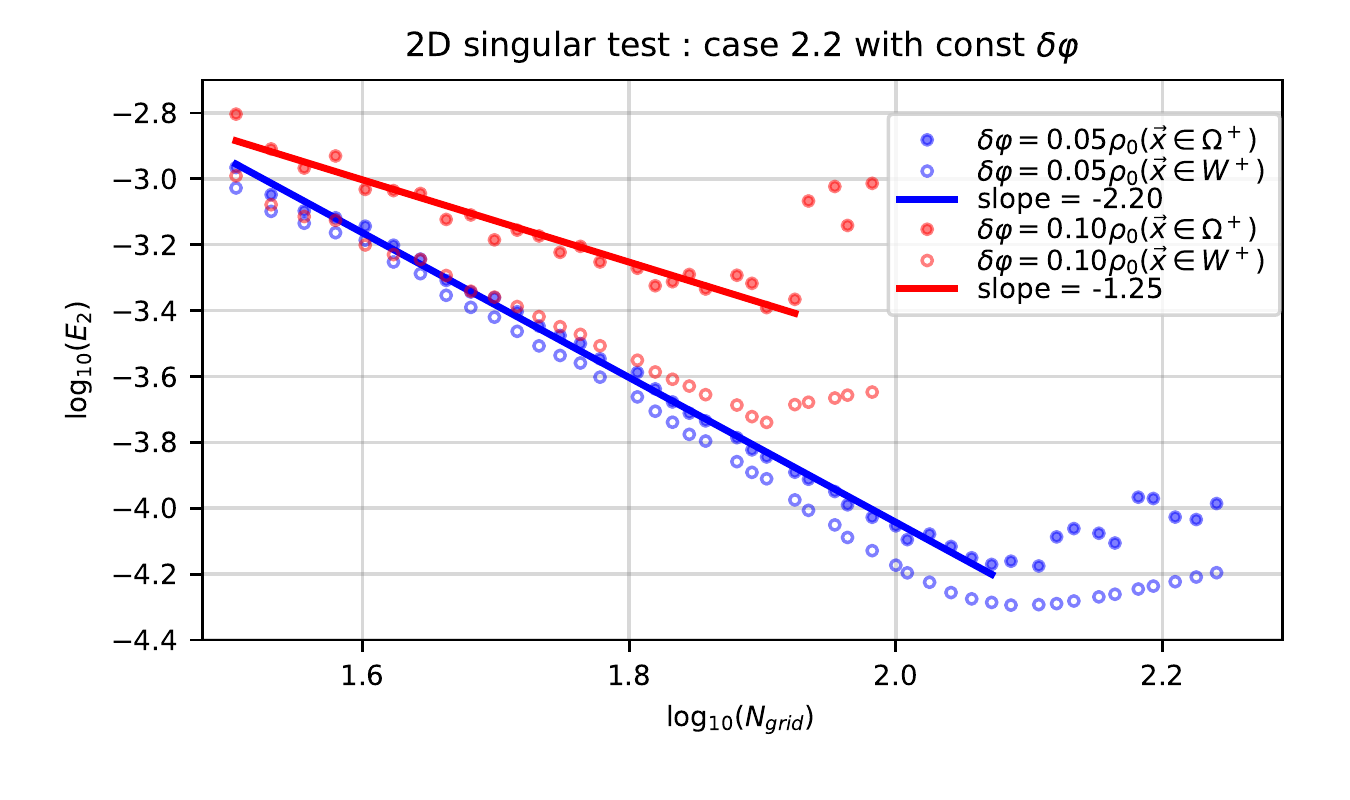}
    \caption{Convergence plot for 2D singular test cases with constant separation $\delta \varphi$. The top plot shows the convergence of 
    test case~$2.1$; the bottom plot shows the convergence of test case~$2.2$. The unfilled points are the $L_2$ error of the result directly calculated from the source term method, which is only defined in $W^+$; the color-filled points are the error of the extrapolated result to the full domain $\Omega^+$. In both test cases, the solver shows convergence up to some resolution, and after that the convergence is lost and the error starts to grow for higher resolutions.}
    \label{fig:plot_2_1&2}
\end{figure}

\begin{figure}[!htbp]
    \centering
    \includegraphics{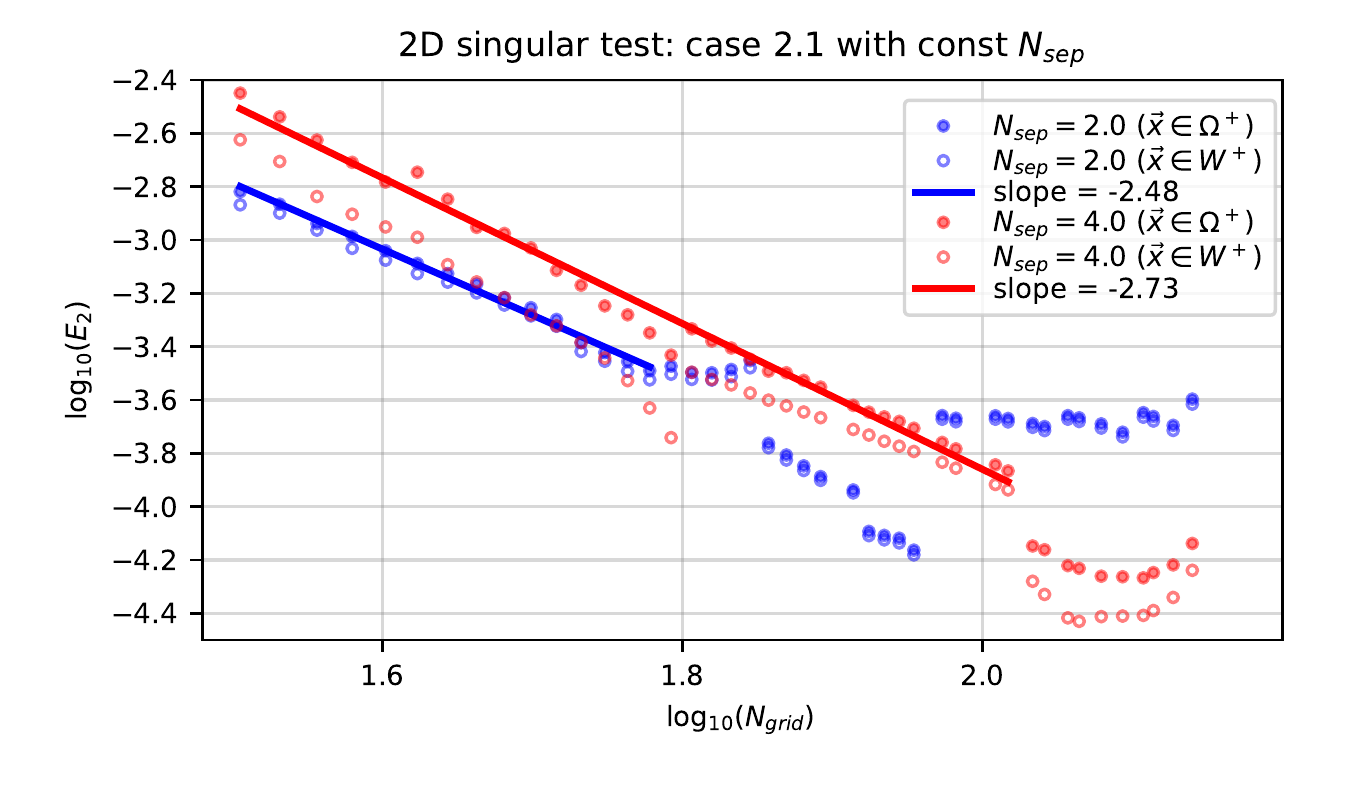}
    \caption{Convergence plot for 2D singular test case with constant separation number $N_{sep}$. The plot presents two sets of separation values with $\delta\varphi = 2.0\Delta x, 4.0 \Delta x$. The unfilled points are the error of the result directly calculated from the source term method, which is only defined in $W^+$; the color-filled points are the error of the extrapolated result to the full domain $\Omega^+$. Similar to Fig.~\ref{fig:plot_2_1&2}, the solver shows convergence up to some resolution. However, as the resolution increases from this threshold, the error stays at approximately the same value as opposed to the increasing error for the constant separation case in Fig.~\ref{fig:plot_2_1&2}.}
    \label{fig:plot_2-3}
\end{figure}

Here, two convergence tests with singular coefficients are performed: a) constant separation $\delta \varphi$ (which implies $\delta r$ is also a constant) with both cases~$2.1$ and $2.2$, where the convergence plots are presented in Fig.~\ref{fig:plot_2_1&2}; b) constant separation number $N_{sep}$ with case~$2.1$, where the convergence plot is presented in Fig.~\ref{fig:plot_2-3}.
For the singular coefficient tests with constant separation $\delta \varphi = 0.05\rho_0$, $0.1 \rho_0$ in Fig.~\ref{fig:plot_2_1&2}, the $L_2$ error defined in $\Omega^+$ shows case~$2.1$ in the top panel with a convergence rate of 
$\sim 1.58$ up to $N_{grid} = 162$ for $\delta \varphi = 0.05 \rho_0$ and a convergence rate of $\sim 2.86$ up 
to $N_{grid} = 180$ for $\delta \varphi = 0.1 \rho_0$. In the bottom panel it shows
case~$2.2$ with a convergence rate of $\sim 2.2$ up to $N_{grid} = 122$ for $\delta \varphi = 0.05 \rho_0$ 
and convergence rate of $\sim 1.25$ up to $N_{grid} = 86$ for $\delta \varphi = 0.1 \rho_0$.

The $L_2$ error of the singular test cases defined in the domain $\Omega^+$ (as indicated by the color-filled points in Fig.~\ref{fig:plot_2_1&2}) shows similar patterns of convergence as the non-singular test. Furthermore, beyond the resolution where the error defined in $\Omega^+$ ceases to show monotonic convergence, the error increases with the resolution. Fig.~\ref{fig:plot_2_1&2} shows that this turn-over point occurs at smaller resolutions for larger separations. The increase in the error is incurred by the extrapolation of the result from $W^+$ to $\Omega^+$. Beyond the turn-over point, the extrapolation error dominates the total error as the resolution increases (with fixed $\delta r$, $N_{sep}$ increases when $\Delta x$ decreases, and larger $N_{sep}$ results in larger extrapolation error).

To further inspect the convergence with singular test cases, we fix the separation number $N_{sep}$ as a constant instead of the separation $\delta\varphi$ in the second set of singular tests. Here we use $N_{sep} = 2.0, 4.0$ with case~$2.1$, where the convergence plot is shown in Fig.~\ref{fig:plot_2-3}. The result of the error defined in $\Omega^+$ (the color-filled points) shows a convergence rate of $\sim 2.48$ up to $N_{grid} = 62$ for $N_{sep} = 2.0$ and a convergence rate of $\sim 2.73$ up to $N_{grid} = 108$ for $N_{sep} = 4.0$. 
Beyond this resolution, the achieved accuracy stagnates but the solution does not show a significant increase in error, as opposed to the singular tests with constant separation.

\begin{figure}[!htbp]
    \centering
    \includegraphics{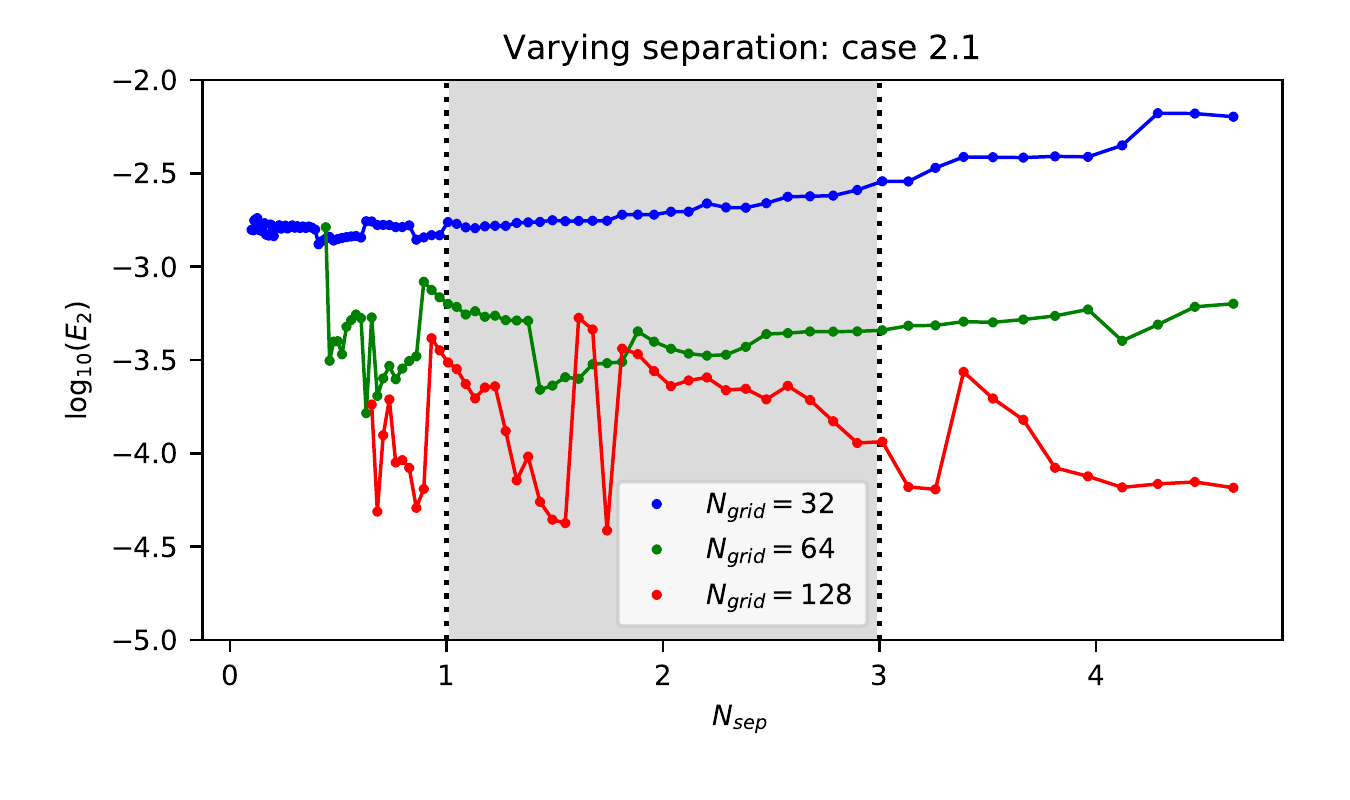}
    \caption{The plot shows the error for test case~$2.1$ at resolutions $N_{grid} = 32, 64, 128$ with various separation number $N_{sep}$ in the x-axis. 
     The two vertical lines indicate the recommended upper and lower limits for $N_{sep}$: a) a lower limit at $N_{sep} = 1.0$ where the error grows sharply for $N_{sep} < 1.0$, which is expected since $\rho \rightarrow 0$ when $N_{sep} \rightarrow 0$ ; b) an upper limit at $N_{sep} = 3.0$ since the extrapolation error dominates the final error for higher separation number. Given these two issues, the recommended value for $N_{sep}$ is between $1.0$ and $3.0$.}
    \label{fig:plot_2_sigma}
\end{figure}

The result in both test cases illustrates that a carefully-chosen separation $\delta \varphi$ in the level set can solve the issue of having an instability in the source term method solver. To further explore the region of suitable separation values, Fig.~\ref{fig:plot_2_sigma} shows a numerical experiment with test case~
$2.1$ at resolutions $N_{grid} = 32, 64, 128$ where the separation number is varied $N_{sep} \in [0.32, 4.82]$.

As indicated by Fig.~\ref{fig:plot_2_sigma}, a suitable value for $N_{sep}$ is approximately between $1.0$ and $3.0$. However, for a general case where the level set is given as tabulated values $\varphi_{i,j,k}$ instead of an analytical function $\varphi({\vec{x}})$, it would not be possible to analytically compute the separation $\delta \varphi$ that satisfies $1.0 \Delta x < \delta r < 3.0 \Delta x$. For such cases, it is recommended to take $W^+ = \Omega^+ / N_2$ (note that $N_2$ here is a neighboring set of $\Omega^+$) such that two grid points are taken as the separation in all directions. As an example, Table~\ref{tab:test_singular} provides five different test cases with singular coefficients, and the treatment $W^+ = \Omega^+ / N_2$ is performed with $N_{grid} = 128$:
\begin{table}[!htbp]
    \centering
    \begin{tabular}{|c | c c c | c c c|}
         \hline
         No. & $\Phi_{theory}$      & $\rho$                      & $\zeta$
        &  $E_{2}$        & iterations & $\delta \varphi$ \\ \hline
        3.1 &  $y+\sin{x}$       & $r^2_0 - r^2$      & $\rho$ 
                & $1.011 \times 10^{-4}$ & $80$  & $ 0.0728 \rho_0$\\
        3.2 &  $y+\sin{x}$       & $r_0 - r$      & $\rho$ 
                & $2.885 \times 10^{-4}$ & $73$  & $0.0371 \rho_0$\\
        3.3 &  $y+\sin{x}$       & $\sqrt{r_0} - \sqrt{r}$      & $\rho$ 
                & $3.086 \times 10^{-4}$ & $76$  & $ 0.0187 \rho_0$\\
        3.4 & $1+\cos{x}\sin{y}$ & $r_0^2 - \bar{r}^2$ &  $\rho$
                &  $2.925\times 10^{-5}$ & $72$    & $0.0874  \rho_0$  \\
        3.5 &  $y+\sin{x}$       & $r_0 - r$      & $\rho e^{r}$ 
                & $3.251 \times 10^{-4}$ & $67$ & $0.0371  \rho_0$  \\
        \hline
    \end{tabular}
    \caption{A table of five singular coefficient test cases with $N_{grid} = 128$ using a separation that gives the domain $W^+ = \Omega^+/N_2$. The $L_2$ error $E_2$, number of iterations, and the separation value are shown for each test. For test case~$3.4$, the modified radius $\bar{r}\equiv \sqrt{(\epsilon x)^2+y^2}$ is used to form an elliptic domain, where the eccentricity $\epsilon = 1.2$; the radius of the domain $r_0=\sqrt{e/4} \approx 0.824$.}
    \label{tab:test_singular}
\end{table}

These two test cases show that the source term method is able to solve 2D Poisson equations that converges with resolution. In general, non-singular cases would show second order convergence since a second-order finite difference Jacobi method is used as the Poisson solver. The non-singular test shown in Fig. \ref{fig:plot_1-1} exhibits symmetry in $x$ and $y$ axes, resulting in the super-convergence that has a convergence rate of $2.5$.
For singular cases, there are multiple contributions of the error other than the Poisson solver, such as the additional separation and extrapolation, and coefficient singularities. Thus, it is difficult to predict the theoretical convergence rate for singular test cases.

\subsection{Irrotational binary neutron stars}
\label{sec:result3}
 This section presents a realistic irrotational BNS test case. The BNS system is set up with two equal mass neutron stars with a polytropic equation of state 
 $P(\rho) = K\rho^\Gamma$ where $\Gamma = 2$, $K = 123.6$, and central density $\rho_0 = 9.574\times 10^{-4}$ ($5.905\times 10^{14}\si{g} \si{cm}^{-3}$). 
 The two stars orbit around each other with an angular orbital velocity $\Omega = 9.075\times 10^{-3}$ ($1.842\times10^{3} \si{rad} \si{s}^{-1}$) 
 at a distance $D_0 = 30.47$ ($45\si{km}$). The parameters listed above are in geometric units.
 
 The initial condition for the source term method solver is generated using the \textsc{cocal} solver ~\cite{Tsokaros_2015}\footnote{We note here that in the \textsc{cocal} solver fluid
 surface fitted coordinates are implemented in order to solve for the fluid potential.}
 in a $301^3$ grid, in which the $60^3$ grid at the center that encapsulates 
 the star is taken as the input for the test case. This includes conformal factor 
 $\psi$, lapse $\alpha$, shift $\beta^i$, density $\rho$, specific enthalpy $h$, and the time 
 component of the four-velocity $u^t$. 
 
 \begin{figure}[!htbp]
    \centering
    \label{fig:plot_3_1}
    \begin{minipage}{.49\textwidth}
        \centering
        \includegraphics[width=\linewidth]{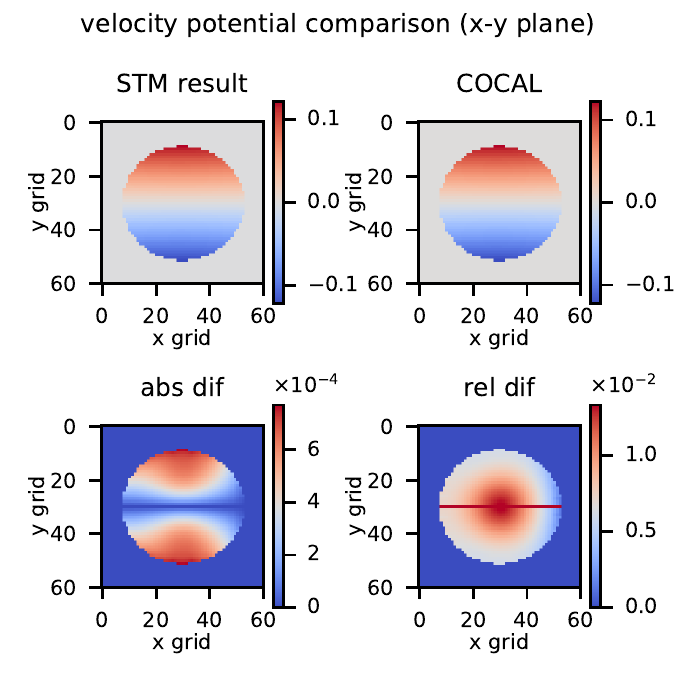}
    \end{minipage}%
    \begin{minipage}{.49\textwidth}
        \centering
        \includegraphics[width=\linewidth]{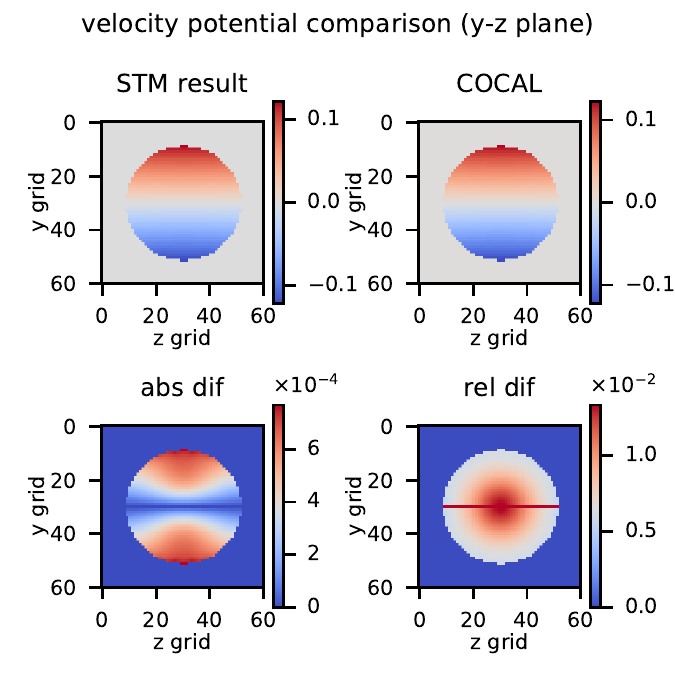}
    \end{minipage}
    \caption{Velocity potential slices in the x-y plane (the left panel) and the y-z plane 
    (the right panel) of the irrotational BNS test case. The two sets of 
    subplots both show the velocity potential $\Phi$ of the source term method (top left), the velocity 
    potential $\Phi_{COCAL}$ generated by \textsc{cocal} (top right), the absolute difference (bottom left) 
    and the relative difference (bottom right) between the two solvers. Regions where the relative difference blows-up have $\Phi_{COCAL}\approx 0$.}
    \label{fig:plot_3}
\end{figure}

\begin{figure}[!ht]
        \centering
        \begin{minipage}{.32\textwidth}
            \centering
            \includegraphics[width=\linewidth]{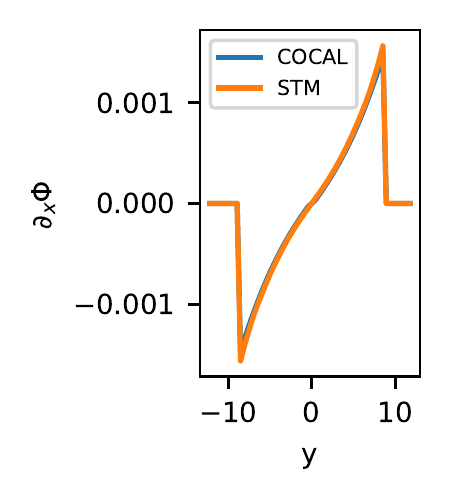}
        \end{minipage}%
        \begin{minipage}{.32\textwidth}
            \centering
            \includegraphics[width=\linewidth]{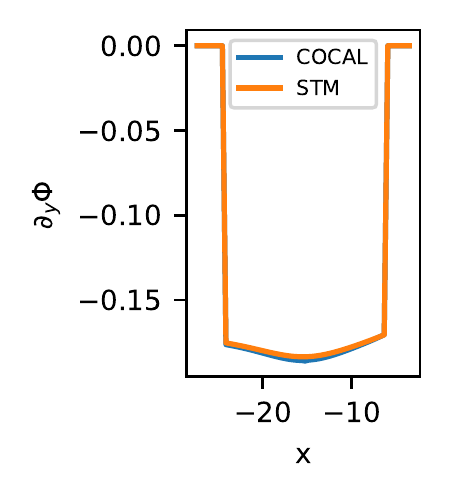}
        \end{minipage}%
        \begin{minipage}{.32\textwidth}
            \centering
            \includegraphics[width=\linewidth]{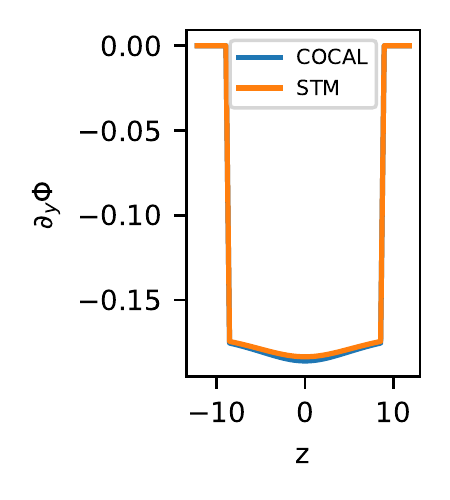}
        \end{minipage}
        \caption{Comparison plots of the gradient of the velocity potential. The left panel shows $\partial_x\Phi$ along the y-axis; the middle panel shows $\partial_y\Phi$ along the x-axis; the right panel shows $\partial_y\Phi$ along the z-axis. It is evident from the plot that the derivative of the velocity potential from the source term method (curves in blue) match well with that from the \textsc{cocal} solver (curves in orange). }
        \label{fig:plot_vel}
\end{figure}

The treatment for the singularity of the coefficients near the surface uses $W^+ = \Omega^+ / N_2$, as described in the previous section. In other words, $\delta\varphi = \min{(\varphi(\vec{x} \in W^+))}$. Figure~\ref{fig:plot_3} shows the comparison between the BNS velocity potential from the source term method and from \textsc{cocal}.
Figure~\ref{fig:plot_vel} shows a comparison of the velocity components, which are the most relevant in
the BNS equations since the potential $\Phi$ never appears in them by itself but only as a gradient. 
The maximum difference obtained in the velocity potential from Fig.~\ref{fig:plot_3} is $\sim 7.7\times10^{-4}$. The maximum relative difference obtained in the velocity potential is $\sim 1.33\times 10^{-2}$ and appears at the center of the grid. Given the fact that the  \textsc{cocal}
code uses spherical coordinates centered at this point its computational error is larger there
\cite{Tsokaros:2016eik}. On the other hand, our computational grid for the source term method is Cartesian with no
coordinate singularity at the center of the NSs. Therefore, we expect our solution to have better 
accuracy at the points close to the center. Overall, this realistic BNS test case demonstrates that the source term method is able to reproduce the velocity potential of the \textsc{cocal} with maximum relative difference no larger 
than $1.4 \% $.

\section{Summary}
To obtain initial data for an irrotational or spinning BNS system, one must solve the conservation of rest-mass
and the Euler equation for a velocity potential. This process results in a Poisson equation with Neumann-type boundary condition on the surface of the stars, and this boundary value problem calls for a robust Poisson solver on an unknown domain.
In this work we apply the source term method~\cite{STM_2018} to solve this type of equation in Cartesian coordinates, 
with the boundary conditions incorporated as jump conditions on the nonlinear sources.

A problem arises for neutron stars (but not with quark stars) near the surface as some of the 
coefficients are approaching zero, causing a divide-by-zero singularity in the source terms. We proposed a solution that introduces a separation $\delta\varphi$ that shifts the level set $\varphi = \rho - \delta \varphi$ to get a smaller domain $W^+$ within the domain of interest $\Omega^+$ such that the coefficients are non-singular inside $W^+$. The solution is then extrapolated from $W^+$ to $\Omega^+$. Given an aptly-chosen separation, the error of the extrapolated solution converges as resolution increases up to some threshold resolution.

Using the source term method, we have implemented a separate elliptic solver in Cartesian coordinates and solved the BNS fluid equation for the velocity potential,
while keeping the gravitational potentials and other thermodynamic variables fixed. We then compared this solution 
with the one coming out from the \textsc{cocal} binary solver and found agreement to the level of $\sim 1\%$.

One advantage of our method is its simplicity. It does 
not require complex surface fitted coordinates and can be implemented in a Cartesian grid which is a standard 
choice in numerical relativity calculations. Another advantage is that now the fluid Poisson equation
can be solved similarly (for example using the Green's function approach \cite{Tsokaros_2015}) to the
gravitational equations simplifying the overall iteration scheme. Finally, we expect that our method can be
used in other problems where non-smooth sources exist, as for example in magnetized rotating neutron stars.
 

\section{Acknowledgements}
This work is funded by the Students Pushing Innovation (SPIN) program in National Center for Supercomputing Applications (NCSA), University of Illinois at Urbana-Champaign.
RH gratefully acknowledges support through NSF grants OAC-2004879, OAC-1550514, and ACI-1238993.
AT is supported by NSF Grants No. PHY-1662211 and No. PHY-2006066, and NASA Grant No. 80NSSC17K0070 to the
University of Illinois at Urbana-Champaign.
This research is part of the Blue Waters sustained-petascale computing project, which is supported by the National Science Foundation (awards OCI-0725070 and ACI-1238993) the State of Illinois, and as of December, 2019, the National Geospatial-Intelligence Agency. Blue Waters is a joint effort of the University of Illinois at Urbana-Champaign and its National Center for Supercomputing Applications.

\section*{References}
\bibliographystyle{iopart-num.bst}
\bibliography{References.bib}

\section*{Appendix}

\subsection{The derivation of the source term method}
\label{appdx:stm_derivation}
Here we present the derivation, as presented in~\cite{STM_2010}, of the source term 
equation (\ref{eqn:source_term}) that we used to solve the fluid boundary value problem.

Consider the domain of Fig.~\ref{fig:domain_1} and the BVP of Eq.~(\ref{eq:gbvp}). 
Integrals over $\Omega^+$ can be converted to integrals over the whole domain $\Omega$  by writing 
\begin{equation}
\int_{\Omega^+} f dV = \int_{\Omega} f H(\varphi) dV ,
\label{eq:vi}
\end{equation}
where $H(z)$ is the Heaviside function
\begin{equation*}
H(z) = 
\cases{
1, & $z \geq 0$, \\
0, & $z < 0$.
}
\end{equation*}
Similarly one can convert integrals over the boundary $\partial\Omega^+$ into volume integrals
over $\Omega$ as follows
\begin{equation}
\int_{\partial\Omega^+} f dS = \int_{\Omega} f \delta(\varphi) |\nabla\varphi| dV ,
\label{eq:si}
\end{equation}
where $\delta(\varphi)=H'(\varphi)$ is the Dirac delta function.

For any sufficiently smooth function $\psi$ one can write 
\begin{equation}
\int_\Omega \psi \nabla^2\Phi dV \ =\ \int_\Omega \Phi\nabla^2\psi dV + \int_{\partial\Omega} \left(\psi\frac{\partial\Phi}{\partial n} - \Phi\frac{\partial\psi}{\partial n}\right) dS.
\label{eq:poiint}
\end{equation}
We will apply the equation above separately in the domains $\Omega^+$ and $\Omega^-$ and then combine them in order
to get an equation in the whole domain $\Omega$. We note that for the $\Omega^+$ case there is only one boundary
$\partial\Omega^+$ while for the $\Omega^-$ case we have two boundary terms on $\partial\Omega^+$ and $\partial\Omega$. We also assume that the test function $\psi$ as well as its derivatives are continuous across
the boundary $\partial\Omega^+$. Using Eqs.~(\ref{eq:vi}), (\ref{eq:si})  and denoting 
$S(\vec{x})=S^+ H(\varphi(\vec{x})) + S^-(1-H(\varphi(\vec{x})))$ 
we obtain
\begin{eqnarray}
\int_\Omega \psi S dV =&  \int_\Omega \Phi\nabla^2\psi dV +
\int_{\partial\Omega} \left(\psi\frac{\partial\Phi}{\partial n} - \Phi\frac{\partial\psi}{\partial n}\right) dS \nonumber\\
&+  \int_{\Omega} \left(\psi a - b\frac{\partial\psi}{\partial n}\right)\delta(\varphi)|\nabla\varphi| dV,
\label{eq:eq1}
\end{eqnarray}
where $a = \Phi_n^+ - \Phi_n^-$ and $b = \Phi^+ - \Phi^-$.
We remove the normal derivative on $\psi$ from the last term above by using integration by parts. It is
\begin{eqnarray}
\int_\Omega \psi b \nabla^2 H(\varphi) dV  &=   -\int_\Omega (b\nabla\psi + \psi\nabla b)\cdot\nabla H(\varphi) dV 
\nonumber \\
&=  \int_\Omega \left(b\frac{\partial\psi}{\partial n} + 
     \psi\frac{\partial b}{\partial n}\right) \delta(\varphi)|\nabla\varphi| dV ,
\label{eq:eq2}
\end{eqnarray}
since the integral over $\partial\Omega$ vanishes there. Using Eq.~(\ref{eq:eq2}) we write Eq.~(\ref{eq:eq1}) as
\begin{eqnarray}
 &\int_\Omega \Phi\nabla^2\psi  dV + 
  \int_{\partial\Omega} \left(\psi\frac{\partial\Phi}{\partial n} - \Phi\frac{\partial\psi}{\partial n}\right) dS
\nonumber \\
&=\int_{\Omega} 
\psi \left( b\nabla^2 H - \left(a +\frac{\partial b}{\partial n}\right)|\nabla\varphi|\delta(\varphi) + S\right) dV 
\nonumber \\
&=\int_{\Omega} 
\psi \left( \nabla^2(bH) - H\nabla^2 b - \left(a -\frac{\partial b}{\partial n}\right)|\nabla\varphi|\delta(\varphi) + S\right) dV.
\qquad\qquad\qquad
\label{eq:eq3}
\end{eqnarray}

Equations~(\ref{eq:poiint}), (\ref{eq:eq3}) imply the source term Eq.~(\ref{eqn:source_term}) in the weak sense.

\subsection{Finite difference discretization of the Heaviside and delta function}
\label{appdx:heaviside_delta_derivation}

This section discusses the finite difference discretization of the Heaviside and delta function in 
Eqs.~(\ref{eq:Heaviside}) and (\ref{eq:Delta}), as demonstrated in sec.~4 of~\cite{STM_2010}. 
The discretization of the Heaviside function is derived in detail in~\cite{Heaviside_2009}, and the 
delta function in~\cite{Delta_2007}.
The goal of the discretization of both functions is to better approximate the integrals
\begin{equation}
    \mathcal{I_H} = \int_\Omega{H(\vec{x})}dV \quad 
    \mathcal{I_\delta} = \int_\Omega{\delta(\vec{x})}dV  .
\end{equation}
Using $I(\varphi)$, $J(\varphi)$, $K(\varphi)$ defined in Eq. (\ref{eq:IJK}), we can write
\begin{eqnarray}
    \nabla^2 I(\varphi) =& \nabla \cdot (H(\varphi)\nabla\varphi) = \delta(\varphi)\abs{\nabla\varphi}^2 + H(\varphi)\nabla^2\varphi , \label{eq:lapI}\\
    \nabla^2 J(\varphi) =& \nabla \cdot (I(\varphi)\nabla\varphi) = H(\varphi)\abs{\nabla\varphi}^2 + I(\varphi)\nabla^2\varphi , \label{eq:lapJ}\\
    \nabla^2 K(\varphi) =& \nabla \cdot (J(\varphi)\nabla\varphi) = I(\varphi)\abs{\nabla\varphi}^2 + J(\varphi)\nabla^2\varphi . \label{eq:lapK}
\end{eqnarray}
Equation~(\ref{eq:lapI}) gives the formula for $\delta(\varphi)$ by substituting $H(\varphi)$ from Eq.~(\ref{eq:lapJ})
\begin{equation}
    \label{eq:delta_derived}
    \delta(\varphi) = \frac{\nabla^2 I}{{\abs{\nabla\varphi}}^2} - \frac{H(\varphi)\nabla^2\varphi}{{\abs{\nabla\varphi}}^2} = \frac{\nabla^2 I}{{\abs{\nabla\varphi}}^2} - \frac{(\nabla^2 J - I\nabla^2\varphi)\nabla^2\varphi}{{\abs{\nabla\varphi}}^4}.
\end{equation}
Similarly, Eq.~(\ref{eq:lapJ}) gives the formula for $H(\varphi)$ by substituting $I(\varphi)$ from 
Eq.~(\ref{eq:lapK}):
\begin{equation}
    \label{eq:H_derived}
    H(\varphi) = \frac{\nabla^2 J}{{\abs{\nabla\varphi}}^2} - \frac{I(\varphi)\nabla^2\varphi}{{\abs{\nabla\varphi}}^2} = \frac{\nabla^2 J}{{\abs{\nabla\varphi}}^2} - \frac{(\nabla^2 K - J\nabla^2\varphi)\nabla^2\varphi}{{\abs{\nabla\varphi}}^4}.
\end{equation}
The expressions for $\delta(\varphi(\vec{x}))$ and $H(\varphi(\vec{x}))$ give the discretization formulas for 
both functions in Eqs.~(\ref{eq:Delta}), (\ref{eq:Heaviside}).

\nocite{LevelSet_book}
\nocite{Dirichlet_2005}
\nocite{FourthLaplace_2005}

\end{document}